\documentclass[twocolumn, twocolappendix]{aastex631}

\shorttitle{A global optical / X-ray correlation of compact objects}
\graphicspath{{./}{figures/}}

\begin{document}

\title{Scaling Up: A Global Optical/X-ray Correlation For Reprocessing Across Black Hole Masses}

\correspondingauthor{Changhyun Cho}
\email{choch@yorku.ca}

\author[0000-0002-9879-1749]{Changhyun Cho}
\affiliation{Center for Astrophysics and Space Science, NYUAD Research Institute, New York University Abu Dhabi, PO Box 129188, Abu Dhabi, UAE}
\affiliation{New York University Abu Dhabi, PO Box 129188, Abu Dhabi, UAE}
\affiliation{New York University, 726 Broadway, New York, NY 10003}
\affiliation{Department of Physics and Astronomy, York University, 4700 Keele Street, Toronto, ON M3J 1P3, Canada}
\author[0000-0002-3500-631X]{David M. Russell}
\affiliation{Center for Astrophysics and Space Science, NYUAD Research Institute, New York University Abu Dhabi, PO Box 129188, Abu Dhabi, UAE}
\affiliation{New York University Abu Dhabi, PO Box 129188, Abu Dhabi, UAE}
\author{Aaryan Sharma}
\affiliation{Center for Astrophysics and Space Science, NYUAD Research Institute, New York University Abu Dhabi, PO Box 129188, Abu Dhabi, UAE}
\affiliation{New York University Abu Dhabi, PO Box 129188, Abu Dhabi, UAE}
\affiliation{Institut Supérieur de l'Aéronautique et de l'Espace (ISAE-SUPAERO), 31055 Toulouse, France}
\author[0000-0002-5319-6620]{Payaswini Saikia}
\affiliation{Department of Astronomy, Yale University, PO Box 208101, New Haven, CT 06520-8101, USA}
\author[0000-0003-1285-4057]{M. Cristina Baglio}
\affiliation{INAF–Osservatorio Astronomico di Brera, Via Bianchi 46, I-23807 Merate LC, Italy}
\author[0000-0003-0168-9906
]{Kevin Alabarta}
\affiliation{INAF–Osservatorio Astronomico di Brera, Via Bianchi 46, I-23807 Merate LC, Italy}
\author[0000-0001-7590-5099]{Sandeep Rout}
\affiliation{Center for Astrophysics and Space Science, NYUAD Research Institute, New York University Abu Dhabi, PO Box 129188, Abu Dhabi, UAE}
\affiliation{New York University Abu Dhabi, PO Box 129188, Abu Dhabi, UAE}
\author[0000-0001-9108-573X]{Yi-Jung Yang}
\affiliation{Center for Astrophysics and Space Science, NYUAD Research Institute, New York University Abu Dhabi, PO Box 129188, Abu Dhabi, UAE}
\affiliation{New York University Abu Dhabi, PO Box 129188, Abu Dhabi, UAE}
\author[0009-0002-6843-7076]{Eleonora Caruso}
\affiliation{Anton Pannekoek Institute for Astronomy, University of Amsterdam, Science Park 904, NL-1098 XH Amsterdam, The Netherlands}
\affiliation{SRON Space Research Institute Netherlands, Niels Bohrweg 4, NL-2333 CA Leiden, The Netherlands}
\author{Mokhine Motsoaledi}
\affiliation{Center for Astrophysics and Space Science, NYUAD Research Institute, New York University Abu Dhabi, PO Box 129188, Abu Dhabi, UAE}
\affiliation{New York University Abu Dhabi, PO Box 129188, Abu Dhabi, UAE}
\author{Meriana Abd Allah}
\affiliation{Center for Astrophysics and Space Science, NYUAD Research Institute, New York University Abu Dhabi, PO Box 129188, Abu Dhabi, UAE}
\affiliation{New York University Abu Dhabi, PO Box 129188, Abu Dhabi, UAE}
\author{Ian McHardy}
\affiliation{Department of Physics and Astronomy, University of Southampton, Southampton SO17 1BJ, UK}

\begin{abstract}
This study extends the optical/X-ray correlation from black hole low-mass X-ray binaries (XRBs) to active galactic nuclei (AGN), bridging scales from stellar-mass to supermassive black holes (SMBHs). In XRBs, optical emission predominantly originates from the thermal reprocessing of high-energy photons on the surface of the geometrically thin, optically thick accretion disk. The reprocessed optical luminosity follows a power-law relation of the form $L_{V} \propto L_{X}^{\beta} a$, with $\beta \sim 0.5$, where $L_X$ is the X-ray luminosity and the orbital separation $a$ serves as a proxy for the disk size. To test the universality of this relationship across all mass scales, we analyzed quasi-simultaneous X-ray and optical flux data from a sample of XRBs and AGN that have estimates of the radius of the optically emitting disk. To bridge the structural differences between these populations, we utilized the characteristic optical/X-ray time lag derived from reverberation mapping as a physical proxy for the reprocessing radius. Our analysis reveals a tight, global correlation across all black hole masses, described by $\log L_V = 0.48\log L_X + 0.96\log a_R + 6.5$ with a scatter of $\sim 0.20$ dex, where $a_R$ is the reprocessing radius derived from the lag. Incorporating a geometric correction for the disk inclination angle ($\cos i$) slightly reduces this scatter to $0.18$ dex. These results suggest that X-ray reprocessing is scale-invariant across accretion disks, from stellar-mass black holes to SMBHs. This scaling relation provides new constraints for models of X-ray emission regions in accreting black holes.
\end{abstract}


\section{Introduction} \label{sec:intro}

Accretion onto black holes appears to scale with mass from stellar-mass X-ray binaries (XRBs) to supermassive black holes (SMBHs) in active galactic nuclei (AGN), as shown by the Fundamental Plane of black hole activity \citep{merloni2003, Falcke2004} and by the mass scaling of X-ray variability timescales \citep{mchardy2006}. Whether the coupling between optical and X-ray emission follows the same scaling has not been established. Given that black holes lack a baryonic surface and are described solely by mass, spin, and charge \citep{israel1967}, it is reasonable to hypothesize that their accretion and outflow processes are governed by similar physical principles. Investigating whether AGN exhibit behavior comparable to stellar-mass black holes when scaled by mass could bridge gaps in our understanding of black hole physics across different regimes. In this work we test one such scaling: the reprocessing of X-ray photons into optical emission on the outer accretion disk, a mechanism that predicts a markedly flatter optical/X-ray dependence than the disk--corona couplings usually invoked to unify these populations.

Efforts to explore scaling relationships between stellar-mass and SMBHs have shown promising results. \citet{merloni2003} and \citet{Falcke2004} demonstrated that these systems align along a ``Fundamental Plane'' of black hole activity, linking black hole mass, X-ray luminosity, and radio luminosity. This result supports the scale-invariance of black hole accretion and jet production across 10 orders of magnitude in mass. Expanding on this, \citet{kording2006} introduced disc-fraction luminosity diagrams (DFLDs) alongside the Fundamental Plane, further highlighting the universal nature of accretion-jet coupling across black hole mass scales. Similarly, \citet{saikia2015} proposed an alternative Fundamental Plane using [O III] emission line luminosity as a bolometric proxy, demonstrating that accretion-jet coupling remains consistent across systems of different sizes. In addition to these luminosity-based studies, \citet{mchardy2006} established a universal scaling relationship between variability timescale, black hole mass, and accretion rate ($T \propto M_{\text{BH}}/\dot{m} $), indicating that black hole variability is governed by shared physical mechanisms. Finally, \citet{koljonen2015} revealed a strong anti-correlation between the X-ray photon index and jet spectral break frequency in both XRBs and AGN, providing evidence that jet formation is tightly linked to plasma conditions near the black hole, independent of mass.

Black hole X-ray binaries (BH XRBs; hereafter simply XRBs, as we do not consider systems with neutron-star accretors) are binary systems comprising a stellar-mass black hole, typically with masses spanning from a few solar masses to $\sim$20 solar masses, and a companion star. Material transferred from the companion forms an accretion disk around the black hole \citep{ss73}. As this material spirals inward, the disk heats up, emitting X-rays from its inner, hotter regions \citep{abramowicz1998theory,abramowicz1998theory2}, while longer-wavelength optical light arises from the cooler, outer regions illuminated by reprocessed X-ray photons \citep{1976ApJ...208..534C}. Variability in X-ray emission causes corresponding changes in the optical light due to the finite light travel time across the disk \citep[see, e.g.,][]{hynes1998}. Measuring these time lags provides valuable insights into the disk's size and geometry \citep{obrien2002}. The presence of the companion star simplifies estimating the disk's dimensions compared to systems without such a companion. During outbursts, XRBs alternate between a hard state, dominated by Comptonized power-law emission from a hot inner flow and accompanied by a compact jet, and a soft state, dominated by thermal disk emission, with intermediate states in between \citep{Fender2004, Remillard2006}.

AGN, with black hole masses ranging from $10^5$ to $10^{10} M_{\odot}$, are at least several orders of magnitude more massive than stellar-mass black holes. Similar to XRBs, AGN exhibit X-ray and optical variability, but on timescales longer by orders of magnitude (seconds in XRBs versus minutes to days in AGN), reflecting the larger size of the system \citep{mchardy2006}. In the reprocessing picture, variations in the X-ray emission drive longer-wavelength variations as the X-rays irradiate the disk, with delays set by the light-travel time across the disk. Reverberation mapping \citep{RN56}, which measures the time lag between emission at different wavelengths, has become an effective tool for estimating the geometry and size of AGN accretion disks \citep[see][for a recent review]{cackett2021}. Time lags in Seyfert 1 AGN have been successfully measured in a growing number of studies \citep[e.g.,][]{RN51, RN41, edelso2019, RN29}.

Similarities in jet morphology and spectral properties across the mass scale reinforce this picture, and combined radio/X-ray studies with black hole mass measurements have been the principal tool for probing accretion--jet coupling \citep{merloni2003, Falcke2004, saikia2015, 2018A&A...616A.152S}.

While radio and X-ray observations effectively probe the jet-corona connection, understanding the energy partition of the accretion flow itself requires examining the coupling between the viscously heated disk and the X-ray emitting corona. For AGN, a non-linear correlation between the optical/UV and X-ray luminosities is well established, reflecting the physical coupling between the viscously heated accretion disk and the X-ray emitting corona. Early studies demonstrated that the X-ray-to-UV flux ratio decreases at higher Eddington ratios \citep{Vignali2003, Strateva2005, Lusso2010}, yielding a general scaling of $L_{\mathrm{X}} \propto L_{\mathrm{UV}}^{\sim 0.6}$ with a scatter of $\sim$ 0.37 dex, meaning the optical/UV emission grows steeper than the X-rays ($L_{\mathrm{UV}} \propto L_{\mathrm{X}}^{\sim 1.6}$). Through rigorous sample selection, the intrinsic scatter of this relation was reduced to $\sim$ 0.21--0.24 dex \citep{Lusso2016}, enabling the use of quasars as cosmological standard candles \citep{Risaliti2019}.

However, the underlying physics governing this energy partition is still a subject of debate. While recent all-sky surveys using eROSITA have explored the foundational $L_{\mathrm{X}} \propto L_{\mathrm{UV}}^{\sim 0.6}$ slope, they present conflicting views on its stability: carefully selected, flux-limited samples suggest a cosmologically invariant relation \citep{Sacchi2025}, whereas Bayesian analyses of massive populations reveal both redshift evolution and a lack of dependence on the Eddington ratio \citep{Chira2026}. This lack of Eddington dependence sits in direct tension with several prevailing analytical accretion flow models \citep[e.g.,][]{Lusso2017, Arcodia2019}.

Extending this disk--corona relation from SMBHs to stellar-mass black holes requires care, because XRB disks are much hotter than AGN disks and their thermal emission peaks in soft X-rays rather than in the UV. The XRB analogue of the UV/X-ray relation therefore compares two components of the X-ray spectrum, the thermal disk emission and the power-law coronal emission. \citet{Arcodia2020} made this comparison for the soft and intermediate states of GX 339$-$4, using the disk emission at 0.2\,keV, roughly the XRB equivalent of 3000\,\AA\ in AGN, and the coronal emission at 2--10\,keV, both normalized by the Eddington luminosity, and compared the result with radiatively efficient AGN. For the full XRB sample, this relation exhibited significant scatter ($\sim 0.43$ dex). This dispersion might be largely driven by the rapid, transient evolution of the inner disk and corona during XRB state transitions. By restricting the XRB sample to sources with photon index and accretion rate distributions comparable to those of the more stable AGN sample, a mass-scaled overlap between the two populations is obtained, with the scatter reduced to $\sim 0.30$–$0.33$ dex. On the theoretical front, models originally developed to explain these XRB state transitions have been successfully adapted to explain the AGN scaling relation. For instance, the jet emitting disk--standard accretion disk (JED-SAD) model, developed for XRBs \citep{Ferreira2006, Marcel2018a, Marcel2018b, Marcel2019} and later applied to AGN \citep{Barnier2023}, utilizes a two-zone radial geometry where the transition radius ($R_{\mathrm{tr}}$) between the outer cold disk and inner hot flow shifts inward as the accretion rate rises. As the inner hot flow is geometrically compressed, it cannot scale its X-ray production as rapidly as the encroaching UV disk, naturally reproducing a scaling relation where intrinsic viscous UV emission rises much more rapidly than the X-ray corona.

While these previous studies provide a framework for unifying compact objects, they fundamentally link the intrinsic viscous dissipation of the inner thermal disk directly to coronal emission, predicting a characteristically steep scaling ($L_{\rm UV} \propto L_{\rm X}^{\sim 1.6}$). However, X-ray thermal reprocessing of the outer disk provides a physically distinct coupling mechanism that predicts a much flatter dependence ($L_{\rm V} \propto L_{\rm X}^{0.5}$) \citep[][]{RN5, Russell2010}. In XRBs, this reprocessing-driven correlation is known to dominate in the hard state, whereas in AGN (such as Seyfert galaxies), evidence for reprocessing is observed across a broad range of accretion rates. In principle, the mechanism is not restricted to a specific accretion state or narrow luminosity regime, offering a distinct path toward scale invariance.

Studies of individual compact-object systems provide strong empirical support for this reprocessing hypothesis. Tight inter-band correlations driven by X-ray illumination are a well-established feature of XRB hard states \citep{RN6, coriat2009, Saikia2022, Yang2025, Bright2025, Saikia2026}, while detailed multi-wavelength variability studies of individual Seyfert 1 galaxies---such as NGC 2617 \citep{fausna2018}, NGC 3516 \citep{noda2016}, NGC 4151 \citep{edelso2017}, NGC 4395 \citep{mchardy2016}, NGC 4593 \citep{mchard2018}, NGC 5548 \citep{RN41}, and NGC 6814 \citep{troyer2016}---confirm that X-ray re-irradiation drives multi-band variability across diverse accretion modes.

The primary objective of this work is to test whether a single optical/X-ray reprocessing relation, $L_{\rm V} \propto a\, L_{\rm X}^{0.5}$, bridges stellar-mass and supermassive black holes. Our sample spans a wide range in accretion rate, covering Eddington-scaled X-ray luminosities of $L_{\rm X}/L_{\rm Edd} \approx 4\times10^{-5}$ to $0.6$ in XRBs and $\approx 6\times10^{-4}$ to $0.05$ in AGN. To this end, we compiled optical/X-ray time lags from published simultaneous light curves, together with simultaneous or quasi-simultaneous X-ray and optical fluxes from which we determined luminosities. At the lower end of this range the inner disk may be truncated, and synchrotron emission from the jet or hot flow can contribute to the optical band, so a single relation is not guaranteed. However, reprocessing in the outer disk is thought to dominate the hard-state optical emission \citep{RN6}, and we include only sources whose optical lags are consistent with disk reprocessing (Section~\ref{sec:obs}). Possible non-thermal contributions are discussed in Section~\ref{sec:dis}.

The theoretical background underlying this study is detailed in Section \ref{sec:theory}, while the observational data and analysis methods are described in Section \ref{sec:obs}. Section \ref{sec:result} presents the results, which are interpreted and discussed in Section \ref{sec:dis}. Finally, Section \ref{sec:con} concludes our findings and highlights potential directions for future research.

\section{Theoretical Background}
\label{sec:theory}

\subsection{Radiative processes in XRBs and AGN}

Numerous studies have shown a positive correlation between X-ray and optical/infrared (OIR) emission in XRBs during outbursts and in AGN \citep[e.g.,][]{1995ApJ...441..786C, chen1997}. This correlation suggests a physical connection between accretion processes and multi-wavelength emission in both XRBs and AGN. Several primary mechanisms have been proposed to explain these correlations.

The first mechanism is the reprocessing of X-rays by the accretion disk (irradiated disk). In this scenario, X-ray photons from the inner accretion flow illuminate the outer disk, where they are absorbed and re-emitted as optical and infrared radiation \citep{RN5}. The resulting emission lags behind X-ray variations by a delay set by the light travel time across the disk. In XRBs, X-rays can also be reprocessed on the irradiated face of the companion star. However, echo mapping of GRO J1655$-$40 locates the dominant reprocessing site in the outer accretion disk, whose thickened rim almost entirely shields the companion from irradiation \citep{hynes1998, obrien2002}. In all five XRBs in our sample, the measured lags are also shorter than the light-crossing time of the binary, consistent with reprocessing in the disk rather than on the companion. We therefore attribute the reprocessed optical emission to the disk.

Another mechanism is intrinsic viscous heating of the disk (non-irradiated disk). In this case, variability in the mass accretion rate leads to changes in the disk's thermal emission, producing optical flux variations independent of X-ray illumination \citep{RN6}. These variations occur on the viscous timescale, which in XRBs is of order days to weeks. The optical flux of GRO J1655$-$40 began rising about six days before its X-ray outburst \citep{Orosz1997}, and a similar delay of about a week was observed in V404 Cyg \citep{Bernardini2016b}. This is far longer than the reprocessing lags of seconds considered here.

Synchrotron emission from the jet is another mechanism contributing to optical and infrared emission. In the hard state of XRBs, compact relativistic jets can produce significant optical and infrared radiation \citep{corbel2002, 2005ApJ...624..295H, kalemci2013, dincer2014, baglio2018}. This emission originates from relativistic electrons accelerated within the jet, and its optical and infrared variability lags the X-rays by only $\sim$0.1--0.2\,s \citep{casella2010, gandhi2017, paice2019}, reflecting the travel time from the inner accretion flow to the region of the jet where this emission is produced.

In AGN, besides the accretion disk, delayed optical and infrared emission also arises from the broad-line region (BLR) and the dusty torus surrounding the disk. The BLR consists of ionized gas clouds located at larger radii than the accretion disk, which reprocess X-ray and ultraviolet radiation into optical emission with time lags corresponding to their light travel distances \citep{2000ApJ...533..631K, bentz2009}. Beyond the BLR, the dusty torus absorbs high-energy radiation and re-emits it in the infrared on even longer timescales \citep{2006ApJ...639...46S, 2020MNRAS.491.4615K}. These reprocessing regions introduce additional lags in AGN light curves, distinct from those associated with the accretion disk.

\subsection{An optical/X-ray correlation between XRBs and AGN}

During outbursts in XRBs, optical radiation is largely attributed to X-ray reprocessing within the accretion disk, though additional contributions may arise from viscous heating in the disk's outer regions and, occasionally, from jets. The concept of an optically thick, geometrically thin accretion disk reprocessing X-rays was initially introduced by \citet{ss73} in the context of stellar-mass black holes. In this model, X-ray photons emitted from the coronal region strike the accretion disk, where they are partially absorbed and reprocessed, resulting in blackbody radiation emitted in the ultraviolet, optical, and infrared spectra. This blackbody radiation dominates the optical emission of XRBs during outbursts. The measured time lags primarily correspond to the light travel time between the X-ray emitter and the reprocessing medium. By analyzing these lags, we can `echo-map' the structure of the system \citep[e.g.,][]{1991ApJ...367L...5H, obrien2002}. \citet{RN5} showed that, if X-ray heating of the disk dominates the optical emission, the optical luminosity scales with both the X-ray luminosity and the accretion disk size, and confirmed this using the absolute $V$ magnitudes of 12 low-mass XRBs, inferring the orbital separation from the orbital period:

\begin{equation}
    L_V \propto a\, L_X^{0.5},
    \label{eq:1}
\end{equation}
where $L_V$, $L_X$, and $a$ represent the optical luminosity, X-ray luminosity, and the orbital separation, respectively. The outer radius of the disk scales with $a$, because the disk extends to a roughly fixed fraction of the compact object's Roche lobe. The exponent of $L_X$ reflects the surface brightness of the accretion disk, $S_\nu \propto T^{4\beta}$: for general $\beta$, the relation is $L_V \propto a^{2-2\beta} L_X^{\beta}$, and in the optical band $\beta \sim 0.5$ \citep{RN5}, which gives the form above.

This relationship has since been extensively tested in stellar-mass black hole systems, providing strong evidence supporting this correlation. \citet{RN6} demonstrated that the observed X-ray and OIR luminosity correlation in hard-state XRBs aligns well with theoretical predictions, with a slope ($\beta$) of approximately 0.6. Their study found that the optical emission of hard-state XRBs is dominated by X-ray reprocessing in the disk, as indicated by its thermal spectrum, with a smaller contribution from the jet. At high luminosity, the jet instead dominates the near-infrared (near-IR) emission. Conversely, during the soft state, both optical and near-IR emission decline, coinciding with the disappearance of the jet.

Just as correlations between X-ray and OIR emission have been observed in XRBs, similar relationships have been found in AGN, though their specific origins differ. AGN exhibit a correlation between X-ray and ultraviolet (UV) luminosities, often attributed to the interaction between the accretion disk and corona. \citet{Arcodia2019} explored this correlation in radiatively efficient AGN and proposed that it arises from an interplay between accretion disk heating and coronal dissipation, with a dependence on accretion state and black hole spin. Another correlation exists between mid-infrared and X-ray luminosities in AGN. \citet{2009A&A...502..457G} found that mid-infrared emission originates from warm dust in the torus surrounding the AGN and serves as a reprocessing signature of the central X-ray emission. Both XRBs and AGN share a fundamental disk-corona interaction that governs their multi-wavelength variability, highlighting the universality of accretion physics across black hole mass scales.

\subsection{Measuring Time Lags: Echo Mapping in XRBs and Continuum Reverberation Mapping in AGN}

Reprocessing lags have been observed in both XRBs and AGN, providing evidence for their presence across different regimes. In XRBs, optical/X-ray lags on the order of seconds have been observed, consistent with light travel time delays caused by X-ray reprocessing in the outer disk \citep[e.g.,][]{obrien2002, RN232, 2020MNRAS.495L..37V}. Similarly, in AGN, optical/X-ray lags driven by reprocessing have been observed on timescales of a few days \citep[e.g.,][]{fausnaugh2016, edelso2017, cackett2021}. While the terminology differs—echo mapping in XRBs and continuum reverberation mapping in AGN—both describe the same physical process, where optical emission from the accretion disk lags behind X-ray fluctuations by a delay set by the light travel time from the X-ray-emitting corona to the optical-emitting region of the disk.

Beyond reprocessing lags, other types of time lags have also been observed in both XRBs and AGN, arising from different physical mechanisms. One example is lags due to viscous propagation, which occur on longer timescales due to accretion rate fluctuations propagating through the disk, rather than the light travel time delays that characterize reprocessing lags. These lags, ranging from minutes to hours in XRBs \citep{2011MNRAS.414L..60U, 2018NewA...58...84A} and months and years in AGN \citep{2003ApJ...584L..53U, arevalo2008, 2009MNRAS.394..427B}, are significantly longer than light travel time delays. In XRBs, these shorter propagation lags arise in the inner disk; the outer, optically emitting disk responds on viscous timescales of days to weeks, as discussed above. Additionally, in XRBs, lags associated with jet emission have been observed, where infrared or radio variations lag behind X-ray fluctuations, reflecting the propagation of disturbances along the jet to its emitting regions \citep{casella2010, Tetarenko2021}.

Reverberation mapping, first introduced by \citet{bahcall1972} and further developed by \citet{RN56}, is a time-resolved technique designed to probe the inner regions of AGN. It works by analyzing time delays in variability across different wavelength bands, offering insights into the size, geometry, and emissivity of the emitting region. The method relies on the assumption that variability in shorter wavelengths is reprocessed and subsequently observed at longer wavelengths, with the delay and smoothing reflecting light travel times within the reprocessing region. This approach has proven crucial for understanding AGN physics, including the accretion disk structure, corona geometry, and surrounding medium \citep{cackett2021}.

In the simple lamppost reprocessing model, X-ray photons from a relativistic corona above the accretion disk irradiate the outer disk, heating it and inducing re-emission at longer wavelengths \citep{haardt1991}. Since the re-emission of absorbed X-rays occurs instantaneously, the time lags between the X-ray and longer wavelength directly trace the light travel delay rather than a physical delay between absorption and re-emission \citep{2005ASPC..330..237H}. Measuring these lags allows us to estimate the size of the UV and optically emitting region of the accretion disk \citep{RN56} and assess whether the relation in Equation \ref{eq:1} extends to the SMBH scale.

The formalism for reverberation mapping is a convolution operation \citep{fausna2018}
\begin{equation}
    f_R (t) = \int f_S (t-\tau) \Psi(\tau) d\tau,
    \label{eq:reber}
\end{equation}
where $f_S$ is the driving signal light curve, $f_R$ is the reverberation light curve, $\tau$ is the time delay (or ``lag''), and $\Psi(\tau)$ is the transfer function.

For AGN, the size of the accretion disk can be theoretically derived using principles of reverberation mapping and the reprocessing model, assuming a standard thin disk structure \citep{fausnaugh2016, edelso2017, edelso2019}. In the reprocessing model, UV/optical variations are primarily driven by a small, variable X-ray-emitting corona located at the center.

The $\tau$-$\lambda$ relationship, based on a standard thin accretion disk model, suggests that the time delay $\tau$ scales as $\lambda^{4/3}$ \citep{2007MNRAS.380..669C}, representing the light travel time from the central source to the region emitting at 1928 \AA \, (W2 band). This scaling remains valid even when the observed lower-energy X-rays are not the primary driving force behind the UV/optical light curves \citep{edelso2019}.

Under these assumptions, the light-crossing time radius $r$ of an annulus emitting at a characteristic wavelength $\lambda$ can be expressed as follows \citep[][]{fausnaugh2016, edelso2017}:

\begin{equation}
    r = c\tau = 0.09 \left( X \frac{\lambda}{1928 \text{\AA}} \right)^{4/3} M^{2/3}_{8} \left( \frac{\dot{m}_{\text{Edd}}}{0.10} \right)^{1/3} \text{ lt-day}
    \label{eq:lamppost}
\end{equation}
where $X$ is a multiplicative scaling factor, accounting for systematic uncertainties in converting the annulus temperature $T$ to the corresponding wavelength $\lambda$ at a characteristic radius $R$, $M_{8}$ represents the black hole mass in units of $10^8 \,\mathrm{M}_\odot$, and $\dot{m}_{\text{Edd}}$ is the Eddington ratio, defined as $L_{\text{bol}} / L_{\text{Edd}}$; the normalization assumes a radiative efficiency of $\eta = 0.1$.

Assuming that the observed wavelength at an annulus of radius $R$ corresponds to the temperature given by Wien's Law, the scaling factor is $X = 4.97$. In contrast, a more realistic approach using a flux-weighted radius yields $X = 2.49$. The flux-weighted estimation is based on the accretion disk's temperature profile, following $T \propto R^{-3/4}$ \citep{ss73}. In both Wien and flux-weighted cases, the disk is assumed to maintain a fixed scale height-to-radius ratio, with heating derived internally from viscous dissipation and externally from coronal X-ray sources positioned above the disk (the lamppost model).

The size of the accretion disk in XRBs can be estimated from the system's orbital parameters, which depends on the mass ratio. For instance, in V404 Cyg, the disk size is constrained to approximately 55\% of the orbital separation \citep{hynes2019}. In contrast, AGN accretion disks can extend to much larger scales, as they are not fed by an orbiting companion star. As a result, direct measurements of AGN disk sizes are not feasible. Instead, the size of the reprocessing region in AGN disks is inferred from the time lag between X-ray and optical emission, a technique similarly applied to XRBs \citep[e.g.,][]{2009MNRAS.392.1106G, 2024NewA..10602109A}.

Extending this relation to AGN is not straightforward, as AGN disks lack the companion-defined outer boundary that sets the disk size in XRBs. We address this by using the measured optical/X-ray time lag as a common estimate of the reprocessing radius, $a_R$, for both populations. This allows us to compile a single dataset spanning stellar-mass to supermassive black holes and to test whether the reprocessing relation holds across this full range.

\section{Observational data} \label{sec:obs}

Our primary goal is to extend the existing XRB relation (Equation \ref{eq:1}) to the AGN scale. To achieve this, we first conducted a comprehensive literature review to identify XRBs and AGN with well-measured optical/X-ray time lags. A key requirement was that the X-ray variability must precede the UV/optical response, as expected from the X-ray reprocessing model. Therefore, any object where the UV/optical peak occurred before the X-ray peak was excluded from our sample.

For XRBs, we considered any positive lag attributed to reprocessing, provided it fell within the range of 1 to 60 seconds. Lags shorter than 1 second were excluded to prevent contamination from jet-related variability, which in the hard state produces optical and infrared lags of only $\sim$0.1--0.2\,s \citep{casella2010, gandhi2017, paice2019}. Lags longer than 60 seconds were discarded, as such extended delays would either exceed the expected light travel time across the binary system or be inconsistent with the X-ray reprocessing framework. This constraint ensures that the observed time lags could be attributed to X-ray reprocessing, by ruling out contributions from other potential sources of optical variability.

For AGN, we ensured that the observed time lags were aligned with the X-ray reprocessing model rather than alternative sources of variability. Specifically, the time lag follows the scaling relation $\tau \propto \lambda^{4/3}$ predicted for X-ray reprocessing by a standard thin disk \citep{2007MNRAS.380..669C}, where $\tau$ is the time lag and $\lambda$ is the wavelength. This relation characterizes reprocessing-driven delays in the accretion disk as a function of wavelength, with longer wavelengths originating from regions farther out in the disk, where X-ray photons take longer to reach. By applying this selection criterion, we isolated X-ray reprocessing from other potential sources of optical variability.

Finally, we required that all selected targets have simultaneous X-ray and optical flux measurements. When strictly simultaneous observations were unavailable, we included data taken within a 24-hour window. This flux data was used to derive X-ray and optical luminosities, which were then compared with the time lags to verify the optical/X-ray correlation and assess whether the relation observed in XRBs extends to AGN. As a result, our final sample consisted of five XRBs and seven AGN.

The optical/X-ray correlation has been shown to depend on the choice of X-ray energy range, as reprocessing is primarily correlated with the bolometric X-ray luminosity, which itself is influenced by the high-energy X-ray spectrum \citep{bernardini2016}. To account for this, we used bolometric (0.1–250 keV) fluxes or luminosities when available from the literature. For sources without published bolometric values, we derived unabsorbed X-ray luminosities in the 0.1–250 keV range by extrapolating the observed spectra using a simple power-law model with a photon index $\Gamma$ obtained from the literature. The corresponding unabsorbed fluxes were then calculated for all available observations.

When only X-ray count rates were available instead of fluxes (e.g., RXTE All-Sky Monitor [ASM] counts in s$^{-1}$), they were converted to unabsorbed fluxes using the NASA WebPIMMS tool\footnote{\url{https://heasarc.gsfc.nasa.gov/cgi-bin/Tools/w3pimms/w3pimms.pl}}. This conversion was performed using literature values for the hydrogen column density $N_H$ and photon index $\Gamma$, ensuring compatibility across different instruments and observations for the correlation analysis.

We derived dereddened optical luminosities by combining optical fluxes, interstellar reddening, and distances, all sourced from the literature. For each source, distance, optical extinction $A_V$, and neutral hydrogen column density $N_H$ from HI absorption measurements were taken from the literature to convert observed magnitudes or fluxes into luminosities. Optical magnitudes in the $V$, $g^\prime$, $R$, and $r^\prime$ bands were converted to flux densities using the corresponding zero points (Vega for $V$ and $R$, AB for $g^\prime$ and $r^\prime$) and dereddened using the extinction law of \citet{cardel1989}, with the most reliable $A_V$ estimate for each source. Where more than one band was available, the dereddened flux densities were averaged, and the optical luminosity was computed as $L_V = 4\pi d^2\,\Delta\nu\,\langle F_\nu \rangle$, where $\Delta\nu = 1.6\times10^{14}$\,Hz is the frequency range spanned by the $g^\prime$ to $R$ bands (477--641\,nm).

To estimate the radius of the optical emission region in the accretion disk, we use the time lag between X-ray and optical emission. This lag arises from the reprocessing of X-rays into optical photons in the disk, providing a direct measure of the optically emitting region. Since the lag corresponds to the reprocessing site in both XRB and AGN accretion disks, it serves as a proxy for the size of the optical emission region, making it the most direct method for testing the optical/X-ray correlation. This approach is particularly valuable for AGN, where direct measurements of the optical emission region are not possible. For consistency, we apply the same method to XRBs.

\subsection{XRBs} \label{subsec:xrb}

For this study, we analyzed data from five XRBs: GRO J1655$-$40, GX 339$-$4, XTE J1118+480, \textit{Swift} J1753.5$-$0127, and MAXI J1820+070. The properties of the five XRBs are summarized in Table \ref{tab:xrbs}. All X-ray luminosities reported in this work are calculated for the 0.1–250 keV energy range. For all other XRBs, either lags from reprocessing have not been constrained, or quasi-simultaneous optical and bolometric X-ray luminosities were not available.

\begin{rotate}
\begin{deluxetable*}{ccccccccc}
\tablenum{1}
\tablecaption{Properties and data collected for the five XRBs \label{tab:xrbs}}
\tablewidth{0pt}
\tablehead{
\colhead{Source} & \colhead{Distance/kpc} & \colhead{Period/hour} &
\colhead{$M_{co}$/$M_\odot$} & \colhead{$M_{cs}$/$M_\odot$} & \colhead{$A_V$} & \colhead{Time lag/sec} & Inclination/deg & \colhead{Number of} \\
\colhead{name} & \colhead{(ref)} & \colhead{(ref)} &
\colhead{(ref)} & \colhead{(ref)} & \colhead{(ref)} & \colhead{(ref)} & \colhead{(ref)} & \colhead{data}
}
\startdata
GRO J1655$-$40 & 3.2 $\pm$ 0.5 & $62.9 \pm 0.0048$ & 5.4 $\pm$ 0.3 & 1.45 $\pm$ 0.35 & 3.7 $\pm$ 0.3 & 19.3 $\pm$ 2.2 & 70.2$\pm 1.9$ & 5 \\
 & (1) & (1) & (1) & (1) & (2) & (3) & (4)  \\
GX 339$-$4 & 10.5 $\pm$ 4.5 & 42.1 $\pm$ 0.01 & 5.9 $\pm$ 3.6 & 1.06 $\pm$ 0.65 & 3.25 $\pm$ 0.25& 10 $\pm$ 4 & 57.5 $\pm$ 20.5 & 49 \\
 & (1) & (1) & (1) & (1) & (5) & (6) & (1) \\
XTE J1118+480 & 1.7 $\pm$ 0.1 & 4.08 & 7.3 $\pm$ 0.7 & 0.18 $\pm$ 0.07 & $0.053^{+0.027}_{-0.016}$ & $3.45^{+1.55}_{-2.45}$ & 75 $\pm$ 7 & 53  \\
 & (1) & (1) & (1) & (1) & (7) & (8) & (1) \\
\textit{Swift} J1753.5$-$0127 & $3.9 \pm 0.7$ & 3.20 $\pm$ 0.02 & 8.8 $\pm$ 1.3 & 0.20 $\pm$ 0.06 & 1.4 $\pm$ 0.3  & 1.0 $\pm$ 0.5  & 79 $\pm$ 5 & 6  \\
 & (9) & (9) & (9) & (9) & (10) &  (11) & (9)  \\
 MAXI J1820+070 & 2.96 $\pm$ 0.33 & 16.45176 $\pm$ 0.00024 & 7.04 $\pm$ 1.31 & 0.53 $\pm$ 0.25 & 0.558 & 4.1 $\pm$ 1.6 & 73.5 $\pm$ 7.5 & 3 \\
  & (12) & (13) & (14) & (14) & (15) & (16) & (14) & \\
\enddata
\tablecomments{The columns list: source names, distance estimates, orbital periods, compact object masses ($M_\odot$), companion star masses ($M_\odot$), interstellar extinction in the $V$ band ($A_V$), optical time lags with respect to X-rays, inclination angles, and the number of optical/X-ray matched data points used in this analysis. References: (1) \cite{RN225} and references therein; (2) \cite{RN229}; (3) \cite{obrien2002}; (4) \cite{greene2001}; (5) \cite{RN231}; (6) \cite{RN232}; (7) \cite{2003MNRAS.346..689C}; (8) \cite{RN233}; (9) \cite{2025AA...694A.119Y}; (10) \cite{fronin2014}; (11) \cite{hynes2009}; (12) \cite{2020MNRAS.493L..81A}; (13)\cite{2019ApJ...882L..21T}; (14) \cite{2020ApJ...893L..37T}; (15) \cite{2018ApJ...867L...9T}; (16) \cite{2019MNRAS.488L..18K}.}
\end{deluxetable*}
\end{rotate}

\subsubsection{GRO J1655$-$40}

GRO J1655$-$40, discovered in 1994 by the Burst and Transient Source Experiment (BATSE) aboard the Compton Gamma Ray Observatory \citep[CGRO; ][]{harmon1995}, is classified as a ``microquasar'' due to its production of relativistic jets. The distance, orbital period, and masses of the black hole and companion star were adopted from \citet{RN225} and references therein. Notably, \cite{obrien2002} reported an optical/X-ray time lag of 19.3 $\pm$ 2.2 seconds, which was attributed to X-ray reprocessing on the accretion disk.

We used unabsorbed X-ray fluxes from Rossi X-ray Timing Explorer \citep[RXTE; ][]{dunn2010} and optical $V$-band magnitudes from \cite{RN68}, observed with the Small and Moderate Aperture Research Telescope System \citep[SMARTS;][]{Subasavage2010} consortium telescope during the hard state. The optical magnitudes were converted to fluxes and dereddened using the optical extinction value from \cite{RN229}. In total, five simultaneous data points were collected between February 21, 2005, and September 29, 2005, for both bands.

\subsubsection{GX 339$-$4}

GX 339$-$4 was first detected in 1972 by the MIT X-ray detector aboard the Orbiting Solar Observatory 7 (OSO 7) satellite \citep{marker1973}. It is one of the most extensively studied black hole XRBs, known for its recurrent outbursts, its significant variability, and its transitions between spectral states. The distance, orbital period, masses of the black hole and companion star, and inclination were adopted from \citet{RN225} and references therein. The optical/X-ray reprocessing time lag for this system is reported as 10 $\pm$ 4 seconds \citep{RN232}.

For this analysis, we used unabsorbed X-ray fluxes from \cite{dunn2010} and optical $V$-band magnitudes obtained with the SMARTS 1.3-meter telescope \citep{RN65}. The optical magnitudes were converted into fluxes and corrected for extinction based on the reddening values from \cite{RN231}. This process yielded 49 data points across the X-ray and optical bands, covering the observational period from May 5, 2006, to July 1, 2009.

\subsubsection{XTE J1118+480}

Discovered in 2000 by the RXTE satellite's ASM \citep{remill2000}, XTE J1118+480 has drawn significant attention due to its proximity to Earth, low X-ray luminosity, and unusual location in the Galactic halo. The distance, orbital period, masses of the black hole and companion star, and inclination were adopted from \citet{RN225} and references therein. Notably, the system exhibits exceptionally low optical extinction, $A_V = 0.053^{+0.027}_{-0.016}$ \citep{2003MNRAS.346..689C}, and an optical/X-ray reprocessing lag of $3.45^{+1.55}_{-2.45}$\,s. Although the rapid UV/optical continuum variability of this source is dominated by synchrotron emission, \citet{RN233} detected a marginal lagged response in the UV emission lines (N\,\textsc{v}, Si\,\textsc{iv}, He\,\textsc{ii}) relative to the continuum, consistent with reprocessing in the disk. We adopt the center of the 0--6.9\,s lag range expected for disk reprocessing, with an uncertainty spanning the 1--5\,s range over which the N\,\textsc{v} response is observed (their figure 16).

Unabsorbed X-ray fluxes for this study were sourced from \cite{dunn2010}, while optical B- and $V$-band magnitudes were obtained from the Liverpool Telescope \citep[LT; ][]{RN66}. These optical magnitudes were converted into fluxes and dereddened using the extinction value reported in \cite{2003MNRAS.346..689C}. This process resulted in 53 quasi-simultaneous X-ray and optical data points spanning March 29, 2000, to February 3, 2005.

\subsubsection{\textit{Swift} J1753.5$-$0127}

\textit{Swift} J1753.5$-$0127 \citep{palmer2005} was discovered during an outburst in 2005 by \textit{Swift}/BAT \citep{Barthelmy2005}. This system attracted significant interest due to its exceptionally prolonged outburst lasting approximately 12 years, concluding with a subsequent mini-outburst \citep{plotki2016, shaw2018, 2019ApJ...876....5Z}. General properties of the system, including its distance, orbital period, masses of the black hole and companion star, and inclination, have been summarized by \citet{2025AA...694A.119Y} and references therein. The optical/X-ray reprocessing lag for this source, measured by \citet{hynes2009}, is 1.0 $\pm$ 0.5 seconds, though this measurement exhibits slightly greater uncertainty compared to other similar systems.

Six X-ray and optical flux pairs were compiled for this source. One pair, from 2014 April, was taken from \citet{tomsic2015}, with optical fluxes calculated by averaging $V$- and $g^\prime$-band magnitudes recorded between April 2 and 8, and unabsorbed X-ray fluxes from \textit{Swift}, Suzaku, and NuSTAR observations. A second pair, from 2005 August, was taken from \citet{cadoll2007}, combining \textit{INTEGRAL} and \textit{Swift} X-ray fluxes with $V$- and $R$-band magnitudes from the NTT at La Silla Observatory. The remaining four pairs come from \textit{INTEGRAL}/ISGRI and RXTE/ASM data, averaged over \textit{INTEGRAL} observing windows in 2005, 2010, and 2011, together with $V$-band magnitudes averaged over the same windows from \citet{hynes2009} and \citet{cadoll2007} for 2005 and from Las Cumbres Observatory monitoring for 2010 and 2011. All optical fluxes for the hard state period, spanning August 11, 2005, to August 28, 2011, were corrected for dust extinction using reddening values from \citet{fronin2014}.

\subsubsection{MAXI J1820+070}

MAXI J1820+070 (also known as ASASSN-18ey) is a black hole low-mass X-ray binary system discovered during its rising outburst phase in early March 2018 by the All-sky Automated Survey for SuperNovae (ASAS-SN) \citep{2018ApJ...867L...9T} and subsequently observed by XMM-Newton \citep{2019MNRAS.488L..18K}. Key observational parameters of this system, such as its distance ($2.96 \pm 0.33$ kpc), orbital period ($\sim 0.685$ days), black hole mass ($7.04 \pm 1.31\,M_\odot$), companion star mass ($0.53 \pm 0.25\,M_\odot$), and inclination ($73.5^\circ\pm7.5$), were provided by \citet{2019ApJ...882L..21T}, \citet{2020MNRAS.493L..81A}, and \citet{2020ApJ...893L..37T}. The optical/X-ray cross-correlation function (CCF) exhibits a positive lag of $4.1 \pm 1.6$ seconds, supporting the interpretation that the optical emission results from the reprocessing of X-rays \citep{2019MNRAS.488L..18K}.

We retrieved the already unabsorbed, bolometric X-ray fluxes from Table 2 of \citet{2021ApJ...909L...9Z}. These measurements include four epochs between March 13, 2018 and April 16, 2018. Matching optical fluxes for these X-ray epochs were found in \citet{2019AN....340..278R}; however, only three epochs had corresponding optical observations. These optical data were collected near the beginning of the outburst, in the hard state, from observations by the Las Cumbres Observatory and Al Sadeem Observatory. To correct the optical fluxes for interstellar extinction, we applied a $V$-band extinction of $A_V = 0.558$ from \citet{2018ApJ...867L...9T}. Consequently, our analysis is based on these three matched data points.

\subsection{AGN}

Seven AGN data sets were included in this study, with their properties summarized in Table \ref{tab:agns}. Intrinsic optical and X-ray luminosities were calculated following the method described in Section \ref{sec:obs}. For NGC 4395, NGC 4593, and NGC 6814, for which we used published light curves, we selected flux pairs near the extremes of the observed variability (and, for NGC 6814, also its average) in order to span the widest available range in luminosity. Note that all AGN in this study are type 1 Seyferts, including intermediate types (Seyfert 1.5--1.9). Because the X-ray corona and inner accretion disk in many Seyferts are only partially obscured along our line of sight, continuum reverberation lags can be measured directly. The optical/X-ray lag measures the light-travel distance between the centroids of the X-ray and optical emission, which may be somewhat smaller than the full extent of the reprocessing region. Because it is measured in the same way for every source, and AGN with broadly similar structures should have lags that scale in the same way with disk size, it nonetheless provides a consistent size measure across the sample.

Lag measurements in AGN are complicated by reprocessing in components beyond the accretion disk (Section~\ref{sec:theory}): diffuse continuum emission from the BLR can lengthen the optical lags, particularly around the Balmer jump \citep{cackett2021, mchard2018}, and the X-ray/UV lags of many AGN exceed the light-crossing time expected from standard reprocessing models, as discussed below. We therefore applied the following criteria to confirm that the lags of each target are consistent with disk reprocessing.

First, under the standard reprocessing model, the X-ray flux peak precedes the optical flux peak. Additionally, the time lag as a function of wavelength is expected to follow the relation given by \cite{edelso2017}:

\begin{equation} \label{eq:tau}
    \tau = \tau_0 \left[\left( \frac{\lambda}{\lambda_0} \right)^{\frac{4}{3}}-1 \right],
\end{equation}
where $\lambda_0$ is the wavelength of the reference band and $\tau_0$ is the fitted lag between wavelength zero and $\lambda_0$. In the literature, the \textit{Swift}/UVOT uvw2 band is widely used for the reference wavelength $\lambda_0$.

Some AGN exhibit optical/X-ray lags approximately three times larger than expected from the standard thin disk model \citep{shappee2014, mchardy2014}; examples include NGC 5548 \citep{fausnaugh2016}, NGC 4593 \citep{cackett2018}, and Mrk 142 \citep{cackett2020}. Despite these discrepancies, the wavelength-dependent lags generally remain consistent with reprocessing models (Eq. \ref{eq:tau}), making the measured lags reliable proxies for accretion disk sizes. The observed excess in lag primarily arises in the UV/X-ray measurements, while the UV/optical lags follow the expected $\lambda^{4/3}$ dependence, though with a larger normalization than predicted \citep{fausnaugh2016}. In this study, we use the observed optical/X-ray lags directly, except for NGC 4151, for which \citet{edelso2017} attribute the excess X-ray/UV lag to a second reprocessing region rather than to the disk; NGC 4151 also shows the largest X-ray/UV lag excess among the AGN compared by \citet{mchard2018}. For this case, we instead extrapolate the UV/optical lag to estimate the corresponding X-ray lag. Possible explanations for the excess UV/X-ray lag are discussed in Section \ref{sec:dis}.

\begin{deluxetable*}{ccccccccc}
\tablenum{2}
\tablecaption{Properties and data collected for the seven AGN \label{tab:agns}}
\tablewidth{0pt}
\tablehead{
\colhead{Source} & \colhead{Distance/Mpc} & \colhead{$M_{\text{BH}}/10^6 M_\odot$} & \colhead{$A_V$} & \colhead{$N_{\text{H}}/10^{20}$cm$^{-2}$} & \colhead{$\Gamma$} & \colhead{Time lag/day} & Inclination/deg & \colhead{Number} \\
\colhead{name} & \colhead{(ref)} & \colhead{(ref)} &  \colhead{(ref)} & \colhead{(ref)} & \colhead{(ref)} & \colhead{(ref)} & \colhead{(ref)} & \colhead{of data}
}
\startdata
NGC 2617 & 67.10 $\pm$ 4.71 & $32.4^{+63}_{-21}$ & 0.092 & 3.64 & 1.69 & $2.47^{+0.88}_{-0.81}$ & 43 & 118 \\
& (1) & (1) & (2) & (3) & (4) & (1) & (1) &  \\
NGC 3516 & 41.3 $\pm$ 32 & $31.7^{+2.8}_{-4.2}$ & 0.114 & 4.08 & 1.75 & $2.6^{+0.91}_{-0.78}$*& 38  &  7 \\
& (5) & (6) & (2) & (7) & (5) & (5) & (8) &  \\
NGC 4051 & $16.6 \pm 0.3$ & $1.73^{+0.55}_{-0.52}$ & 0.036 &  1.35 & 2.19 & $1.2^{+1.0}_{-0.3}$ & 12 & 1 \\
& (9) & (6) & (2) & (7) & (10) & (11) & (12) &  \\
NGC 4151 & $19.0^{+2.4}_{-2.6}$ & $54 \pm 18$ & 0.074 & 1030$^\dagger$ & 1.67 & $1.30^{+0.48}_{-0.47}$** & 45 & 302  \\
& (13) & (13) & (2) & (14) & (14) & (15) & (12) &  \\
NGC 4395 & $4.3 \pm 0.4$ & $0.36 \pm 0.11$ & 0.4 & 10$^\dagger$ & 1.6 & 9.1$\times10^{-3}$*** & 37 & 3 \\
& (16) & (17) & (18) & (19) & (19) & (20) &  (21) &  \\
NGC 4593 & 35 & $7.63 \pm 0.16 $ &  0.217 & 400$^\dagger$ & 1.84 & 0.91 $\pm$ 0.41  & 30 & 2 \\
& (22) & (22) & (23) & (22) & (22) & (22) & (24) &  \\
NGC 6814 & $21.65 \pm 0.41$ & $18.5 \pm 3.5$ & 0.51 & 3.4$^\dagger$ & 1.53 & $1.9^{+1.1}_{-1.2}$ & 47 & 3  \\
& (25) & (26) & (25) & (27) & (27) & (28) & (29) \\
\enddata
\tablecomments{The columns include: source names, distance estimates, SMBH masses ($10^6 M_\odot$), interstellar extinction in the $V$ band ($A_V$), hydrogen column density ($10^{20} \text{cm}^{-2}$; values marked $^\dagger$ are, or include, absorption intrinsic to the AGN, all others are Galactic), power-law photon index ($\Gamma$), optical ($V$-band) time lags with respect to X-rays, and the number of data points used in this analysis. References: (1) \cite{fausna2018}; (2) \cite{schlaf2011}; (3) \cite{kalberla2005}; (4) \cite{shappee2014}; (5) *extrapolated from X-ray/B-band lag; \cite{noda2016}; (6) \cite{denney2010}; (7) \cite{dickey1990}; (8) \cite{Wu2001}; (9) \cite{yuan2021}; (10) \cite{turner2017}; (11) \cite{breedt2010}; (12) \cite{fischer2013}; (13) \cite{honig2014}; (14) \cite{hinkle2021}; (15) **fitted X-ray/UV lag + observed UV/optical lag; \cite{edelso2017}; (16) \cite{thim2004}; (17) \cite{peters2005}; (18) \cite{lira1999}; (19) \cite{cameron2012}; (20) ***$788^{+44}_{-54}$ s; \cite{mchardy2016}; (21) \cite{brok2015}; (22) \cite{mchard2018}; (23) \cite{clabel1983}; (24) \cite{lu2000}; (25) \cite{bentz2019}; (26) \cite{bentz2009}; (27) \cite{walton2013}; (28) \cite{troyer2016}; (29) \cite{Pancoast2014}.}
\end{deluxetable*}

\subsubsection{NGC 2617}

NGC 2617 is classified as a Seyfert Type 1 AGN, with key observational parameters, including X-ray and optical fluxes, obtained from \citet{fausna2018} and related works (Table \ref{tab:agns}). The X-ray to optical time lag in NGC 2617 is approximately 2.5 days \citep{fausna2018}, derived from the centroid distributions using the Flux Randomization and Random Subset Selection (FR/RSS) method. Notably, no long-term variations were removed from the light curves prior to the analysis. Although this lag is larger than the value predicted by standard reprocessing models by a factor of 2.3 \citep{fausna2018}, a significant correlation remains between the X-ray and UV/optical light curves. Additionally, the lag-wavelength relation closely follows the predictions for reprocessing in a geometrically thin, optically thick accretion disk, justifying the inclusion of NGC 2617 in our analysis.

Note that the optical \textit{Swift} $V$-band flux was corrected for both Galactic extinction and host-galaxy starlight. The \textit{Swift}/UVOT light curves were obtained using standard aperture photometry with a 5\arcsec\ radius circular aperture centered on the AGN and background counts estimated from a 15\arcsec\ surrounding annulus \citep{fausna2018}. The 0.3–10 keV X-ray flux was corrected for absorption by \citet{fausna2018}. To calculate X-ray bolometric luminosity, we used the column density value 3.64$\times 10^{20}$ cm$^{-2}$ from \cite{kalberla2005} and the power-law photon index 1.69 from \cite{shappee2014}. In total, 118 data points were collected between May 1, 2013, and April 8, 2014.

\subsubsection{NGC 3516}

NGC 3516 is a Seyfert Type 1 galaxy, with most of the parameters used in this study adopted from \cite{noda2016}. A correlation between X-ray and B-band flux was detected for the first time in this source, with changes in the hard X-ray component preceding those in the B-band by $2.0^{+0.7}_{-0.6}$ days \citep{noda2016}. To ensure consistency in our analysis, we extrapolated this X-ray/B-band lag to the X-ray/V-band range using the reprocessing scaling relation (Eq. \ref{eq:tau}), resulting in a lag of 2.6 days. The measured optical lag strongly supports the X-ray reprocessing model, suggesting that the optical brightening was driven by irradiation from the hard X-ray intensity \citep[e.g.,][]{1991ApJ...371..541K, 2007MNRAS.380..669C}.

\citet{noda2016} also provided seven time-averaged, background-subtracted X-ray fluxes observed by the Suzaku telescope, alongside the corresponding optical fluxes measured between April 10, 2013, and April 8, 2014. By applying a power-law index of 1.75 \citep{noda2016} and a hydrogen column density of 4.08$\times 10^{20}$ cm$^{-2}$ \citep{dickey1990}, we calculated the intrinsic bolometric X-ray luminosity and the optical luminosity with the host-galaxy contribution removed.

\subsubsection{NGC 4051}

NGC 4051 is a nearby ($z$ = 0.0023) narrow-line Seyfert 1 galaxy \citep{mchardy2004}, and the key observational parameters are presented in Table \ref{tab:agns}.

To obtain the optical luminosity for NGC 4051, we used the time-averaged, dereddened V- and R-band fluxes from \cite{breedt2010}, which were corrected for dust extinction and host galaxy contributions. The $V$-band fluxes were measured from January 8, 2001, to December 11, 2008, while the R-band fluxes were measured from December 28, 2001, to July 28, 2008. The unabsorbed X-ray luminosity was computed from the time-averaged, unabsorbed X-ray flux reported by \cite{breedt2010}, measured between April 13, 1996, and December 11, 2008.

The CCF between the optical light curve and X-rays shows the strongest peak where the optical variations lag behind the X-rays by 1.2$^{+1.0}_{-0.3}$ days \citep{breedt2010}. This time delay is consistent with the light travel time to the optical-emitting region of the accretion disk, indicating that X-ray reprocessing is responsible for the rapid optical variations. The time lags observed between other optical bands and X-rays also support this interpretation. However, a flat-disk reprocessing model cannot account for all of the optical variability. \citet{breedt2010} find a second correlation peak at an optical delay of $39^{+2.7}_{-8.4}$ days, consistent with the dust sublimation radius, and attribute the additional optical flux to the dusty torus, through reprocessing of X-rays by hot dust and reflection of optical light from the disk. Because this component is delayed by weeks, it is well separated from the 1.2-day disk reprocessing lag adopted here, so the data are suitable for our analysis.

\subsubsection{NGC 4151}

NGC 4151 is one of the brightest Seyfert 1 galaxies, with key observational properties summarized in Table \ref{tab:agns}. While this object does not fully adhere to the typical lamp-post reprocessing model, it does follow an emission-reflection relationship ($\tau \sim \lambda^{4/3}$) and suggests that photons from the corona may pass through an additional layer, making it suitable for our study \citep{edelso2017}.

The observed time lag between the X-ray and optical bands is approximately 4.4 days—comprising a $3.39^{+0.33}_{-0.28}$ day lag between the X-ray and UV/optical bands and a $0.96^{+0.47}_{-0.46}$ day lag within the UV/optical bands. This time lag is significantly larger than the expected size of the inner accretion disk or coronal region, suggesting that it is unlikely to originate from the standard reprocessing model. They may instead be associated with a process governed by the dynamical timescale ($t_{\text{dyn}}$), which dictates how the disk responds to the loss of hydrostatic equilibrium \citep{gardner2017}. In this scenario, an extreme-ultraviolet (EUV) toroidal-shaped component prevents the X-ray corona from directly illuminating the accretion disk. Instead, the corona heats the EUV component, which then thermalizes and illuminates the disk. Consequently, we decided to exclude the observed X-ray and UV time lag and instead extrapolated the time lag using UV and optical data points. This approach yielded a fitted X-ray/UV time lag of $0.34 \pm 0.11$ days. Adding this fitted X-ray/UV lag to the observed UV/optical lag gives a combined lag of $1.30^{+0.48}_{-0.47}$ days, which we used in our analysis.

For the simultaneous X-ray and optical fluxes, we used data from \citet{edelso2017}, with \textit{Swift} observations covering the period from February 20, 2016, to April 27, 2016. In the optical analysis, we obtained the $V$-band fluxes from \citet{edelso2017} and subtracted the host galaxy contributions (details of this process can be found in Appendix \ref{app:host}). Additionally, we expanded the X-ray energy flux range from 0.3–10 keV to 0.1–250 keV. For the analysis, we used a photon index of 1.67 and a column density of $1.03 \times 10^{23}$ cm$^{-2}$ from \cite{hinkle2021}.

\subsubsection{NGC 4395}

NGC 4395 is classified as a Seyfert 1.8/1.9 galaxy \citep{filippenko1989} and is a pure disk galaxy, meaning it lacks a bulge \citep{filippenko2003}. It is notable for its low-mass SMBH, $M_{\rm BH} = (3.6 \pm 1.1)\times10^{5}\,M_\odot$, and its proximity, at a distance of $4.3 \pm 0.4$\,Mpc (Table~\ref{tab:agns}). Mass estimates for the central black hole in NGC 4395 span $\sim 10^4$--$4\times10^5\,M_\odot$ \citep{filippenko2003, peters2005, brok2015, woo2019}, reflecting sensitivity to broad-line profile choices and extrapolation below calibrated mass regimes. We adopt the reverberation mass from \citet{peters2005}, which agrees with independent gas-dynamical estimates \citep{brok2015}. The correlations in Figures~\ref{fig:correlation}--\ref{fig:each_only} do not depend on this choice, since $M_{\rm BH}$ enters only the mass-normalized comparison (Figure~\ref{fig:edd}).

The optical/X-ray g-band time lag for NGC 4395 is reported by \citet{mchardy2016} as $788^{+44}_{-54}$ seconds. The tight correlation between the X-ray and UV/optical light curves supports the interpretation that the UV/optical variability in NGC 4395 is primarily caused by the reprocessing of X-ray photons by the accretion disk. Very fast multiband monitoring with HiPERCAM confirms that the UV/optical lags arise from reprocessing in the disk, with little contribution from the BLR, making NGC 4395 one of the cleanest examples of disk reprocessing among AGN \citep{McHardy2023}. Additionally, this time lag is the shortest among AGN used in this study, and with the smallest SMBH mass, NGC 4395 represents an intermediate case between two distinct regimes.

The X-ray and optical fluxes used in this study were obtained from \citet{cameron2012} using \textit{Swift} observations, covering the period from August 2008 to March 2009. From the light curves presented in their Figure 1, we selected three data points corresponding to the highest, second-highest, and lowest flux values. Since the host galaxy contribution had already been subtracted and the data dereddened, no further corrections were necessary. The optical luminosity was calculated based on the galaxy's distance. For the X-ray fluxes, initially measured in the 2–10 keV range, we extrapolated to the 0.1–250 keV range using the power-law photon index ($\Gamma = 1.6$) and hydrogen column density ($N_{\rm H} = 1.0\times10^{21}$\,cm$^{-2}$) provided by \citet{cameron2012}, enabling us to derive the bolometric X-ray luminosity.

\subsubsection{NGC 4593}

NGC 4593 is a well-studied Seyfert 1 galaxy, notable for its significant variability across X-ray, UV, and optical bands. The optical/X-ray time lag, measured at approximately 0.91 days using \textit{Swift} light curve data \citep{mchard2018}, underscores the interconnection between these emission regions. The observed UV/optical time lag of $0.247 \pm 0.38$ days agrees well with the predictions of the standard accretion disk model. However, the X-ray/UV time lag of $0.66 \pm 0.15$ days exceeds the model expectation (0.090 days) by approximately a factor of seven. Despite this discrepancy, the analysis supports the lamppost model, which still provides a good fit to the \textit{Swift} light curves across all wavelengths, suggesting that the X-ray-to-optical emission delays remain consistent with reprocessing in the accretion disk. Unlike in NGC 4151, \citet{mchard2018} find that the disk-reprocessing response peaks at short lags ($<1$ day) and attribute the excess to an extended low-level tail from surrounding gas; when variations on timescales longer than 5 days are filtered out, the X-ray/UV lag approaches the disk-reprocessing expectation. We therefore use the observed lag for NGC 4593.

We used \textit{Swift} observational data collected from July 13, 2016, to August 6, 2016 \citep{mchard2018}. Two data points were selected during this period, representing the peak and minimum flux levels. Prior to calculating the bolometric X-ray and optical luminosities, necessary corrections such as dereddening and removal of host galaxy contributions were applied.

\subsubsection{NGC 6814}

NGC 6814 is a Seyfert 1.5 galaxy, classified as a nearly face-on, grand-design spiral with a Hubble type of SBc. It is one of the closest broad-line Seyfert galaxies, located at a redshift of 0.0054 \citep{bentz2019}, and has been included in prior reverberation mapping campaigns, such as the LAMP project \citep{bentz2009}.

For this analysis, we utilized X-ray and optical flux data from \citet{troyer2016}. NGC 6814 was observed by \textit{Swift} over a three-month period, from June 8 to September 12, 2012, capturing both X-ray and UV fluxes. Optical data were additionally obtained using the Liverpool Telescope \citep{liverpool}. From the provided light curves, we selected the peak, average, and minimum flux values. Since corrections for host galaxy contributions and reddening were already applied, no further adjustments were necessary.

The time delay between the X-ray and optical $V$-band light curves is reported to be approximately 1.9 days. The observed time lags, variability amplitudes, and the smoothing of longer-wavelength light curves are consistent with a scenario where X-rays irradiate and are reprocessed within the accretion disk, driving UV/optical variability. The lag is, however, several times to an order of magnitude longer than the standard thin-disk prediction (Eq.~\ref{eq:lamppost}), making NGC 6814 a useful test case for reprocessing geometries (Section~\ref{sec:alt}).

\section{Results}
\label{sec:result}

\begin{figure*}
\includegraphics[width=1.0\textwidth]{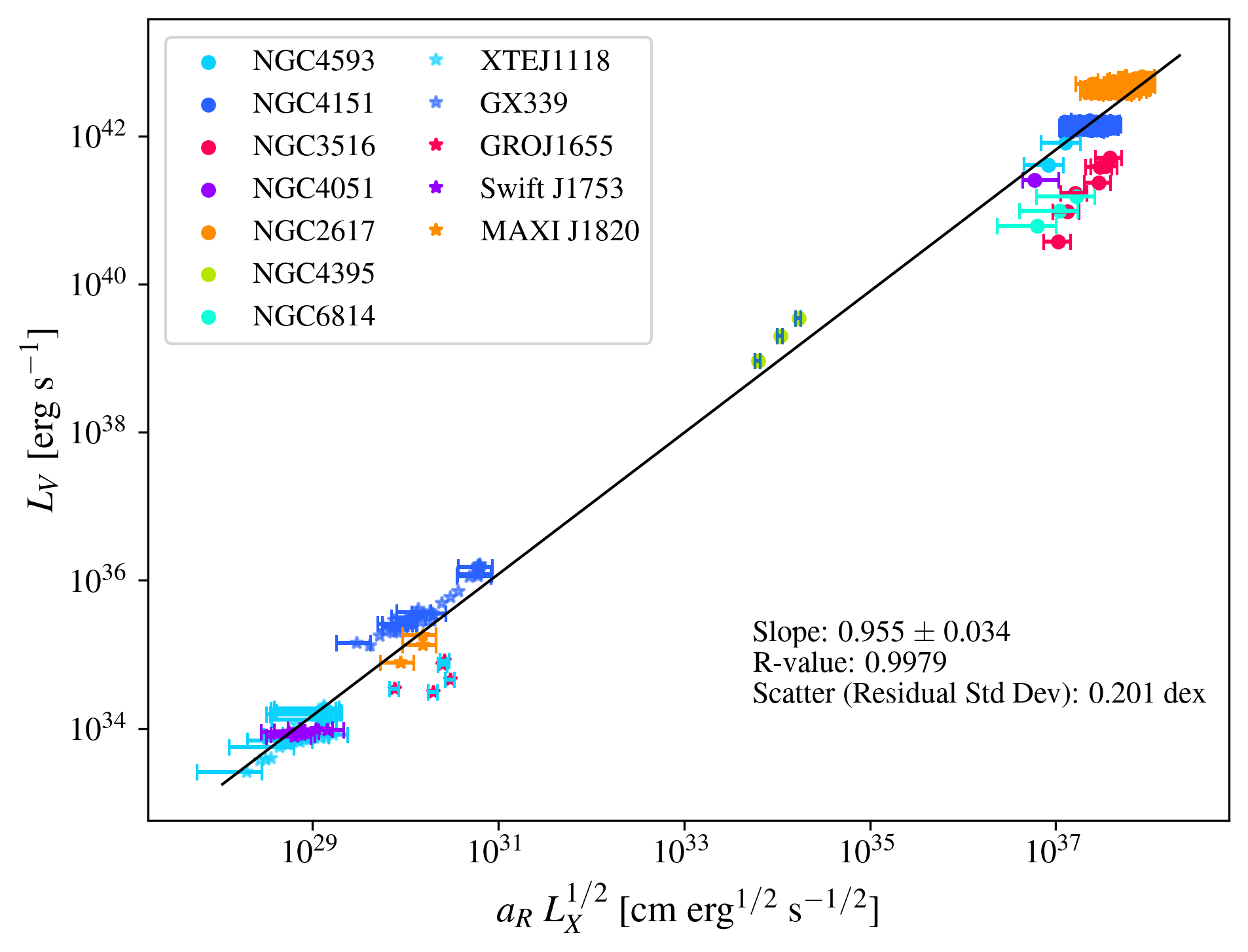}
\caption{The global optical/X-ray correlation for a combined sample of five XRBs and seven AGN. The observed optical luminosity ($L_V$) is plotted against the reprocessing term $a_R L_X^{1/2}$, where $L_X$ is the bolometric X-ray luminosity (0.1--250 keV) and $a_R$ is the reprocessing radius estimated from measured time lags. The solid line represents the best-fit relation (Eq.~\ref{eq:5}).
\label{fig:correlation}}
\end{figure*}

In this analysis, we used five XRBs in the hard state and seven AGN to explore if the existing relation in XRBs can be extended to the AGN scale. Simultaneous X-ray and optical luminosities and the optical/X-ray lags were collected and the host galaxy contributions and the effects of absorptions were all removed.

\subsection{The Global Optical/X-ray Correlation} \label{subsec:global}

Validating Equation \ref{eq:1} on the AGN scale requires addressing the fundamental structural differences between XRBs and AGN. A direct comparison of AGN accretion disk sizes is challenging: an XRB's outer disk radius is typically determined by Roche lobe, shaped by its companion star, whereas AGN disks lack such a boundary and are often partially obscured by the BLR or other extended dusty structures.

We establish a physically equivalent comparison by utilizing the characteristic reprocessing radius of the optical emission ($a_R$). This radius is measured directly from the observed optical/X-ray time lag, which corresponds to the light-travel time from the central X-ray source to the reprocessing site emitting optical photons. This lag-derived radius provides the consistent physical basis required to investigate the $a \sqrt{L_X}$--$L_V$ relationship across both populations.

Figure~\ref{fig:correlation} illustrates the relationship between $L_V$ and $a_R L_X^{0.5}$ for both XRBs and AGN, based on quasi-simultaneous optical and X-ray luminosities. The combined sample spans a range of almost ten orders of magnitude both in optical and X-ray luminosities. The error bars on the data points represent only the uncertainties in the time lag measurements ($a_R$). Formal observational errors on the fluxes are not included: typical photometric uncertainties are $\lesssim$0.03 dex, an order of magnitude smaller than the median lag uncertainty of $\sim$0.17 dex in $\log a_R$, so the lag uncertainty dominates the statistical error budget for all sources except NGC~4395, whose lag is measured to $\sim$6\%. Systematic effects, including the bolometric extrapolation, are discussed in Section~\ref{sec:dis}.

Uncertainties on the fit parameters were estimated by bootstrap resampling of the twelve sources rather than the individual measurements, because the 552 data points are drawn from only twelve independent systems (e.g., NGC~4151 alone contributes 302 points). Least-squares errors that treat every point as independent underestimate the slope uncertainty by a factor of $\sim$13.

We first fit the data with both exponents free, $\log L_V = \beta \log L_X + \gamma \log a_R + c$. This yields $\beta = 0.53 \pm 0.13$ and $\gamma = 0.87 \pm 0.20$, consistent with the value $\beta = 0.5$ expected for X-ray reprocessing (Equation~\ref{eq:1}). The two exponents are strongly anti-correlated ($\rho = -0.95$) because $\log L_X$ and $\log a_R$ are themselves correlated across the sample ($r = 0.99$). We therefore adopt $\beta = 0.5$, which reduces the relation to a one-parameter scaling in $a_R L_X^{0.5}$ (Figure~\ref{fig:correlation}).

With $\beta$ fixed, the best-fit relation yields a slope of $0.955 \pm 0.034$, a Pearson correlation coefficient of $R = 0.998$, and a total scatter of $0.201$ dex, corresponding to an intrinsic scatter of $0.143$ dex once lag uncertainties are accounted for. This reflects a remarkably strong positive relationship across all mass scales, mathematically expressed as:

\begin{equation}
\log L_V = 0.48 \log L_X + 0.96 \log a_R + 6.47,\label{eq:5}
\end{equation}
where $L_V$ and $L_X$ are the optical and X-ray luminosities in units of $\mathrm{erg \, s}^{-1}$, and $a_R = c\tau$ is the reprocessing radius in cm, obtained from the measured optical/X-ray time lag $\tau$. The coefficients of Equation~\ref{eq:5} (0.48 and 0.96) agree with the free-fit exponents within their uncertainties, and removing any single source changes the slope by at most 0.02. Because 76\% of the data points come from NGC~4151 and NGC~2617, we also tested whether these heavily monitored sources dominate the fit. Giving each source equal weight yields a slope of $0.94 \pm 0.03$, and removing both sources together yields $0.89 \pm 0.05$, both consistent with the pooled value.

At the lower-luminosity end of the distribution, the data from all five XRBs align closely with the global trend. Although GRO J1655$-$40 (red star markers) lies systematically below the mean correlation---this deviation may arise partly from its weak jet \citep{RN68} and partly from the different epochs of its lag and flux measurements (Section~\ref{sec:dis})---the overall XRB sample exhibits only minor deviations.

In the high-luminosity end, the AGN sample demonstrates that this scaling relation holds well into the supermassive regime. While the AGN broadly follow the exact same structural track laid out by the XRBs, they exhibit slightly more localized scatter. Specific sources such as NGC 3516 and NGC 6814 fall below the main trend line. Rather than breaking the fundamental correlation, these deviations are probably consequences of complex, source-specific reprocessing geometries---such as truncated disks or inner hot flows---which naturally reduce the solid angle available for direct X-ray irradiation. The viscous disk contribution can also change the position on the correlation, and can make the correlation within an individual AGN flatter (shallower).

Taken together, both XRBs and AGN are consistent with a single relation, $L_V \propto a_R L_X^{0.5}$, over nearly ten orders of magnitude in luminosity, with an intrinsic scatter of 0.143 dex. The best-fit slope is robust to the removal of any individual source, and a fit with free exponents recovers $\beta = 0.53 \pm 0.13$, consistent with the value expected for X-ray reprocessing. This relation describes the sample of sources as a whole; the variability of individual AGN can follow flatter internal slopes, as discussed in Section~\ref{sec:dis}.

\subsection{Inclination-Corrected Scaling and Scatter Reduction}

\begin{figure*}
\includegraphics[width=1.0\textwidth]{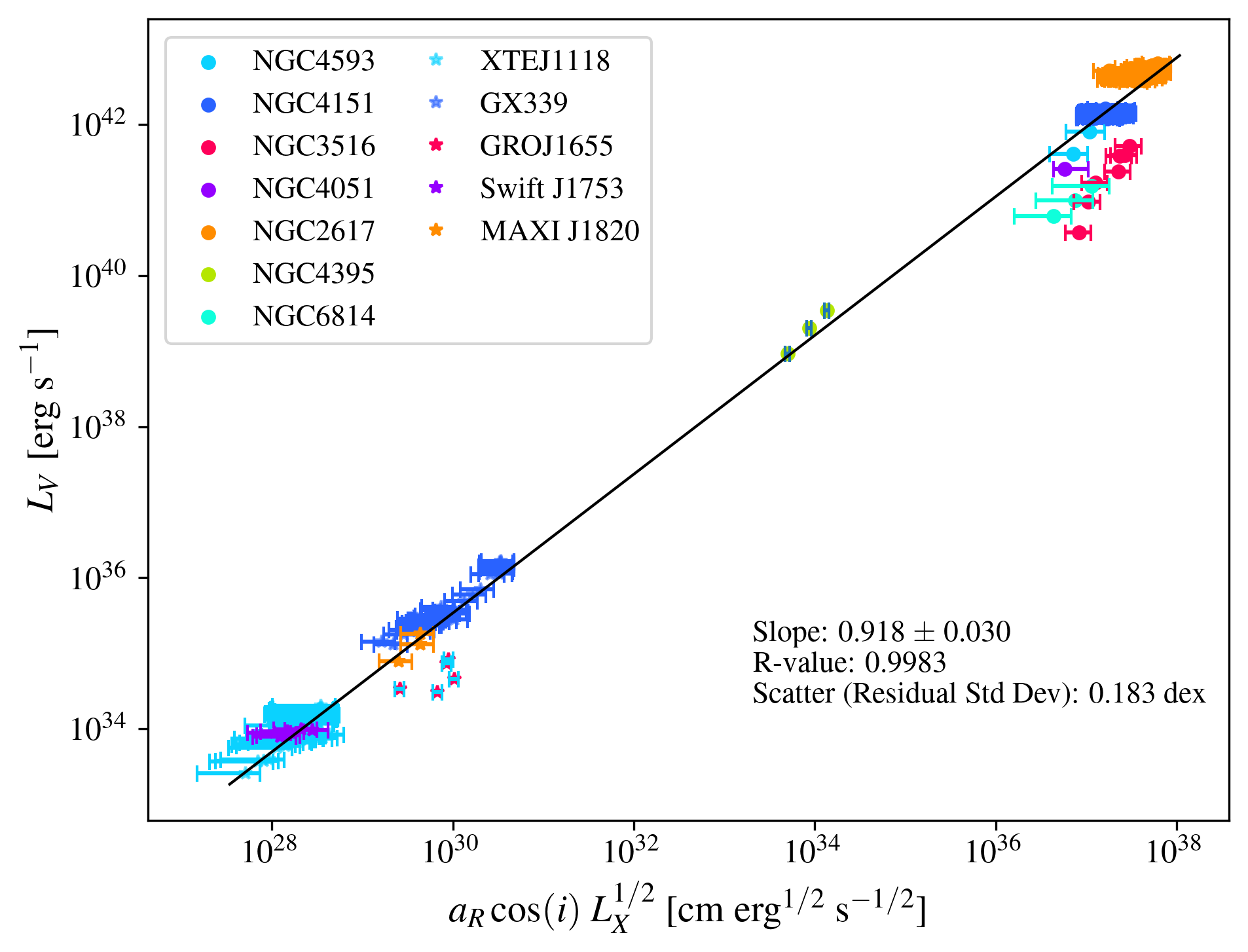}
\caption{The inclination-corrected optical/X-ray correlation for the same sample of XRBs and AGN.}
\label{fig:inclination}
\end{figure*}

Figure~\ref{fig:inclination} extends the analysis by incorporating an inclination correction into the scaling relation, evaluating $L_V$ against $a_R \cos(i) L_X^{0.5}$. This geometric correction is physically motivated because the X-ray and optical emission mechanisms possess different geometries. The primary X-ray emission from the corona is generally assumed to be isotropic, whereas the reprocessed optical emission originates from the surface of a geometrically thin, optically thick accretion disk. Consequently, the observed optical luminosity depends directly on the projected surface area of the disk, so that it depends on the observer's viewing angle.

Applying this inclination correction slightly reduces the scatter of the correlation. Fitting the exponents freely with the inclination correction again gives $\beta = 0.48 \pm 0.12$, consistent with 0.5. With $\beta$ fixed, the best-fit slope is $0.918 \pm 0.030$ (bootstrap over sources; $R = 0.998$), and removing any single source changes it by at most 0.01. The total scatter decreases from $0.201$ to $0.183$\,dex and the intrinsic scatter from $0.143$ to $0.127$\,dex. This is the sense expected if the observed optical luminosity scales with the projected area of the disk. The reduction appears in 75\% of bootstrap realizations, but randomly reassigning the twelve measured $\cos i$ values among the sources yields a scatter at least as small in 10\% of trials, so with twelve independent systems the improvement is suggestive rather than decisive. The AGN inclinations are particularly uncertain, as they are inferred indirectly (e.g., from narrow- or broad-line-region dynamics) rather than from the disk itself. Such errors add scatter, so the improvement found here is, if anything, conservative. This refined correlation is described by the equation:

\begin{equation}
\log L_V = 0.46 \log L_X + 0.92 \log (a_R \cos i) + 7.99,
\end{equation}
where $i$ is the inclination angle. Despite minor deviations in individual slopes, both the XRB and AGN populations appear to align along this inclination-corrected track. The reduction in global scatter upon introducing the $\cos i$ dependence is consistent with the underlying physical model: the optical emission is dominated by thermal reprocessing on a flat disk structure, a mechanism that scales across black hole systems of vastly different masses.

\subsection{Independent Population Analysis}

To test whether the global correlation is merely a statistical artifact of connecting two distinct, widely separated luminosity regimes, we performed independent regressions on the XRB and AGN sub-samples. If the thermal reprocessing mechanism is truly universal, the scaling relations derived from each population in isolation should yield statistically consistent slopes and intercepts.

\begin{figure*}
\centering
\includegraphics[width=1.0\textwidth]{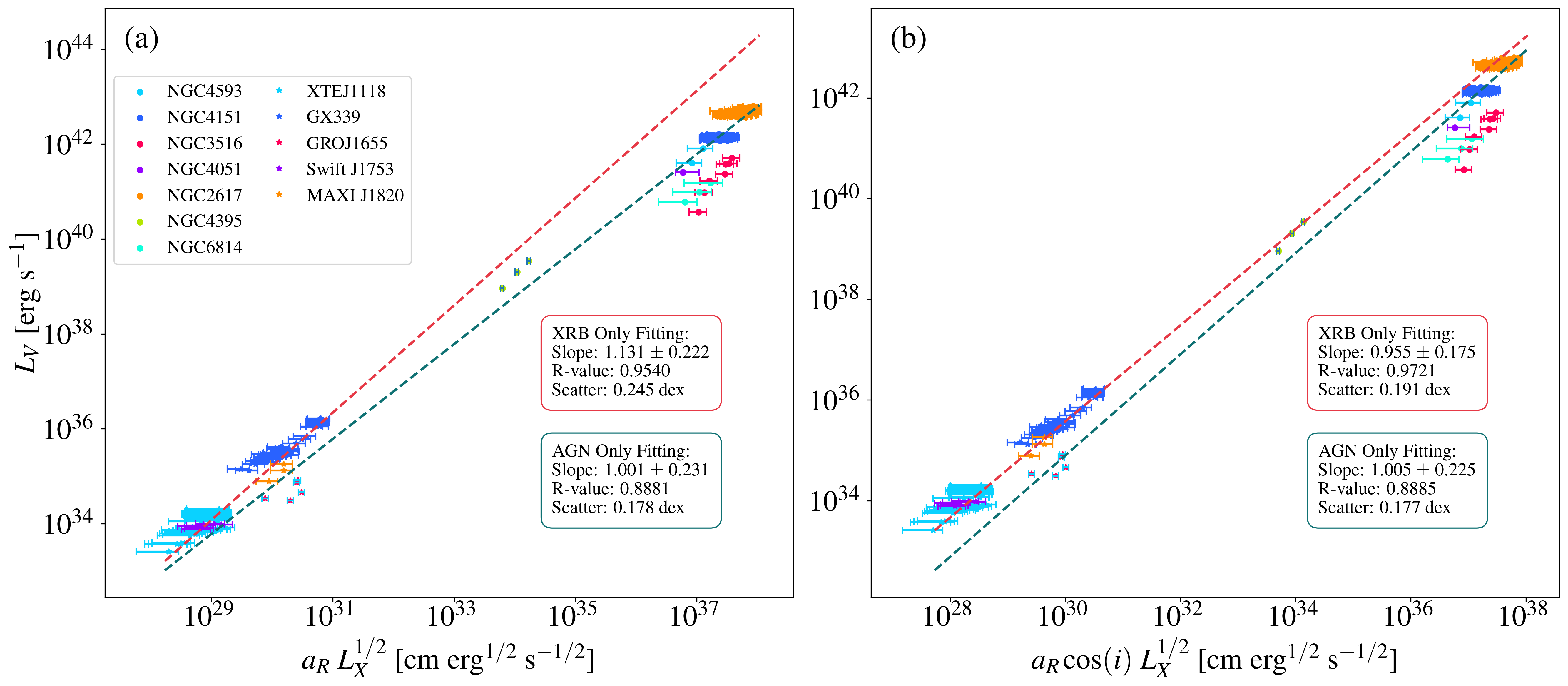}
\caption{Independent regressions for the XRB and AGN populations, demonstrating the underlying consistency of the scaling relations. \textbf{(a)} The uncorrected correlation ($L_V$ vs. $a_R L_X^{0.5}$), showing the individual fit for XRBs (red dashed line) and AGN (green dashed line). \textbf{(b)} The inclination-corrected correlation ($L_V$ vs. $a_R \cos i \ L_X^{0.5}$).}
\label{fig:each_only}
\end{figure*}

Figure~\ref{fig:each_only} presents the results of this independent analysis. The left panel of Figure~\ref{fig:each_only} displays the separated fits prior to any inclination correction. The individual fitting line for the XRB sample (red dashed line) yields a slope of $1.131 \pm 0.222$, while the AGN sample (green dashed line) yields $1.001 \pm 0.231$; both are consistent with the global slope and with each other. Each sub-sample fit is an ordinary least-squares regression on all data points of that population, including sources lying below the relation such as NGC~3516, with uncertainties estimated by bootstrap resampling of the sources. Excluding NGC~3516 changes the AGN slope only from 1.001 to 0.984.

The effect of the accretion disk geometry can be examined further by including viewing-angle effects. As illustrated in the right panel of Figure~\ref{fig:each_only}, applying the $\cos i$ inclination correction brings the two sub-sample slopes into closer formal agreement, although given the size of the sub-samples they were already consistent within their uncertainties before the correction.

In these inclination-corrected versions, separate fits yield a slope of $0.955 \pm 0.175$ for the XRB sample (red dashed line) and $1.005 \pm 0.225$ for the AGN sample (green dashed line). Both independent slopes are consistent with the inclination-corrected global slope of $0.918$ (Figure~\ref{fig:inclination}), indicating that each mass regime is individually consistent with the $L_\mathrm{V} \propto a_R L_\mathrm{X}^{0.5}$ scaling law.

This consistency supports two key physical conclusions. First, the inclination correction improves the consistency of the fits, for both stellar-mass binaries and SMBHs, consistent with the reprocessed optical emission originating from a geometrically thin disk in both systems. Second, because the independent population slopes naturally bridge the mass gap without requiring artificial scaling factors, the thermal reprocessing scenario can account for the optical/X-ray energy distribution across the range of the central black hole mass probed here.

\subsection{Mass-Normalized Optical and X-ray Scaling Relations} \label{subsec:edd}

\begin{figure*}
\centering
\includegraphics[width=1.0\textwidth]{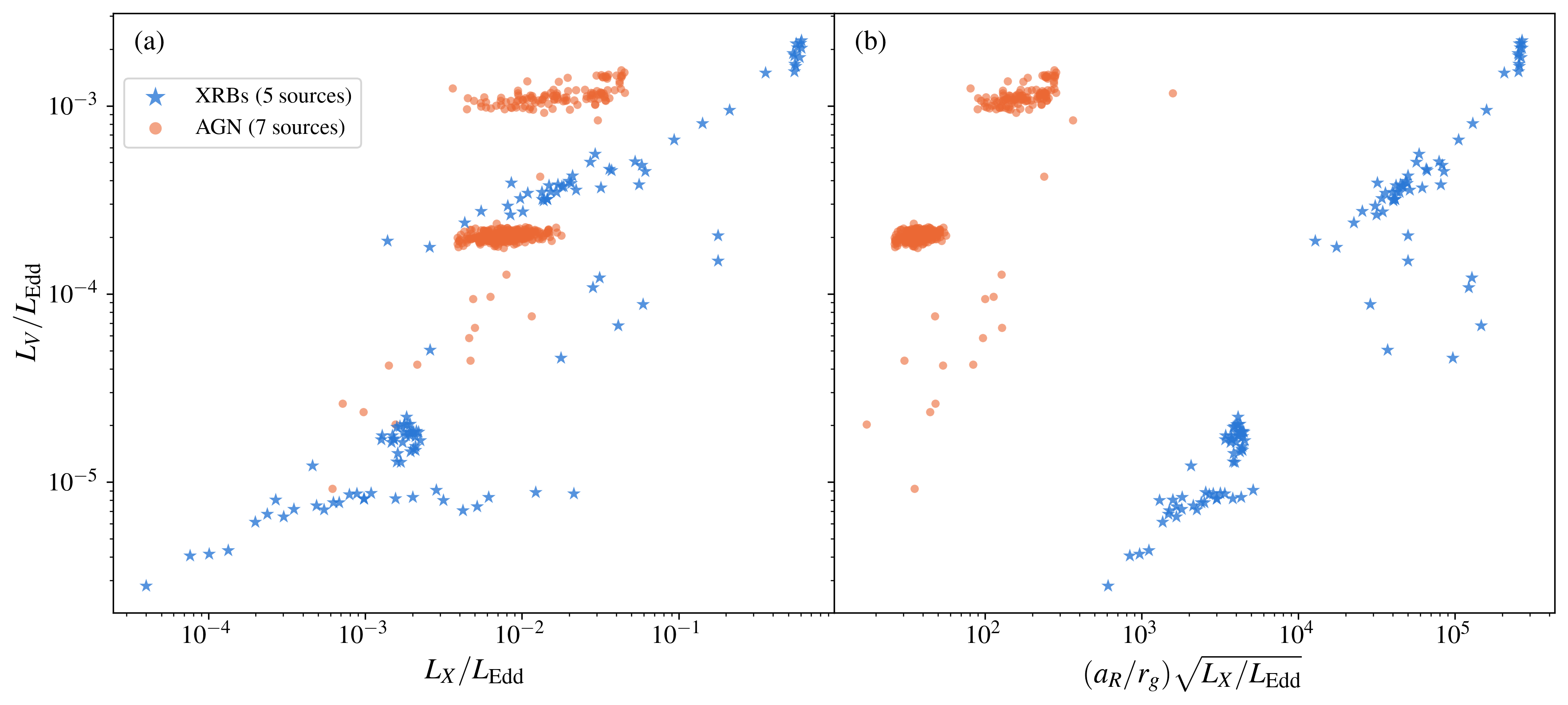}
\caption{Mass-normalized scaling relations for XRBs and AGN. \textbf{(a)} Eddington-scaled optical luminosity ($L_V/L_{\mathrm{Edd}}$) versus Eddington-scaled X-ray luminosity ($L_X/L_{\mathrm{Edd}}$). \textbf{(b)} Eddington-scaled optical luminosity versus the mass-invariant reprocessing parameter, $(a_R/r_g)(L_X/L_{\mathrm{Edd}})^{0.5}$. Blue stars show the five XRBs and orange circles the seven AGN.}
\label{fig:edd}
\end{figure*}

To test whether thermal reprocessing is scale-invariant across the black hole mass spectrum, we construct a mass-normalized parameter space. In Figure~\ref{fig:edd} (a), we plot the optical luminosity against the X-ray luminosity, with both quantities scaled by the Eddington luminosity ($L_{\mathrm{Edd}}$) to account for the large mass difference between XRBs and AGN.

In this Eddington-scaled plane, the two populations occupy a similar region: the AGN range ($L_\mathrm{X}/L_\mathrm{Edd} \approx 6\times10^{-4}$--$0.05$) lies entirely within that of the XRBs ($\approx 4\times10^{-5}$--$0.6$). The selected Seyfert 1 galaxies and hard-state XRBs therefore operate at comparable normalized accretion rates.

Moreover, the Eddington-normalized optical luminosities ($L_\mathrm{V}/L_\mathrm{Edd}$) also span a remarkably similar range across both populations. While some AGN appear slightly brighter in the optical than the general XRB population at a given X-ray Eddington ratio, there is substantial overlap between the two samples. However, only a loose correlation (Pearson $R \approx 0.77$) emerges in this plane, considerably weaker than that in Figure~\ref{fig:correlation}. This indicates that the X-ray Eddington ratio alone is a poor predictor of optical luminosity; rather, explicitly accounting for disk size scaling and X-ray thermal reprocessing is essential to uncover the underlying universal relation. Within the two best-sampled AGN, the relation is much flatter. In Figure~\ref{fig:edd}(a), NGC~2617 and NGC~4151 appear as nearly horizontal clusters, following $L_V \propto L_X^{\beta}$ with $\beta = 0.11 \pm 0.02$ and $0.06 \pm 0.01$, respectively, well below the value of 0.5 expected for pure reprocessing. By comparison, the well-sampled XRBs GX~339$-$4 and XTE~J1118+480 have $\beta = 0.43 \pm 0.02$ and $0.45 \pm 0.03$. Possible causes of this flattening are discussed in Section~\ref{sec:dis}.

To explicitly account for this structural scale, standard reprocessing models predict that the optical emission depends not only on the incident X-ray luminosity but also on the characteristic size of the emitting region, following $L_\mathrm{V} \propto a_R \, L_\mathrm{X}^{0.5}$.

To translate this relation into a scale-invariant framework, we normalize the disk size by the gravitational radius ($r_\mathrm{g}$) and express luminosities relative to the Eddington limit. This yields the physically motivated parameter $(a_R/r_\mathrm{g})\left(L_\mathrm{X}/L_{\mathrm{Edd}}\right)^{0.5}$, which encapsulates the expected scaling of reprocessed optical emission. We use this formulation to test whether scaling by black hole mass is required to reveal the underlying reprocessing relation.

In contrast to the partially overlapping populations in panel (a), AGN and XRBs occupy distinct regions along the horizontal axis in panel (b), reflecting systematic differences in the normalized size of their optical-emitting regions ($a_R/r_g$). This separation arises because the characteristic disk temperature decreases with increasing black hole mass \citep{ss73}, such that optical emission originates at much larger gravitational radii (i.e. in units of $r_g$) in stellar-mass XRBs than in supermassive AGN.

Rather than exhibiting a tight, single trend across the AGN sample in Figure~\ref{fig:edd} (b), the supermassive sources are dispersed across several orders of magnitude. Crucially, this dispersion is expected within the framework of disk reprocessing. The intrinsic reprocessing relation ($L_{\mathrm{V}} \propto a_R \, L_{\mathrm{X}}^{0.5}$) is fundamentally governed by the incident X-ray flux and physical disk radius, independent of black hole mass ($M_{\mathrm{BH}}$). Consequently, normalizing both axes by the Eddington limit introduces an artificial dependence on $M_{\mathrm{BH}}$. Because our AGN sample spans over two orders of magnitude in black hole mass (a factor of $\sim 150$), sources with identical absolute X-ray luminosities and disk dimensions---which emit identical optical reprocessed light---are shifted to vastly different locations in the Eddington-scaled plane. In contrast, the XRB sample exhibits a tighter apparent correlation in this representation primarily because its constituent black hole masses are nearly uniform (spanning only a factor of $\sim 1.6$). The scatter in the Eddington-scaled AGN distribution therefore indicates that black hole mass scaling is not required to uncover the primary reprocessing mechanism, and that absolute X-ray illumination and disk geometry drive the optical emission across all scales.

\section{Discussion} \label{sec:dis}

The results in Section \ref{sec:result} demonstrate a strong correlation between $L_V$ and $a \ L_X^{0.5}$, extending this relationship from XRBs to AGN. However, the observed intrinsic scatter indicates that while the global scaling holds, the underlying populations exhibit source-specific variations. Specifically, certain AGN display individual evolutionary slopes that are shallower than the universal expectation, while a subset of XRBs and AGN resides systematically below the main trend line due to a relative deficit in optical emission compared to their X-ray output (or higher X-ray emission compared to their optical counterpart). To interpret these findings, this section first establishes the theoretical framework of X-ray reprocessing to demonstrate why this mechanism specifically predicts an $L_X^{0.5}$ dependence. We then examine the physical origins of the observed deviations---including X-ray/UV lag excesses and intrinsic disk variability---and discuss the broader implications for accretion physics.

\subsection{Theoretical models of reprocessing}

Theoretical models of X-ray reprocessing (irradiation) and viscous heating (non-irradiation) \citep{RN6, coriat2009, shabhaz2015} suggest that these processes are governed by several key factors: (i) the temperature profile of the accretion disk, where $T \propto R^{-n}$, with $n = 3/4$ for a standard thin disk and $n = 1/2$ for an irradiation-dominated disk; (ii) the surface brightness distribution, $S_\nu \propto T^{4\beta}$; (iii) the regions contributing to optical emission, such as the Rayleigh-Jeans tail, the overlapping/flat region of the spectrum, or the Wien cutoff; (iv) the observed wavelength, which determines the sampled part of the disk; and (v) the relationship between X-ray luminosity and the mass accretion rate, determined by the radiative efficiency. These factors highlight the intricate nature of X-ray/OIR correlations, emphasizing that the dominant emission mechanism varies with the physical conditions of the system.

The region of the disk contributing to optical emission impacts the slope of the surface brightness distribution, $S_\nu \propto T^{4\beta}$, which directly influences the optical/X-ray correlation (Eq. \ref{eq:1}). In the context of XRBs, early theoretical work proposed $\beta = 0.5$ for this correlation \citep{RN5}; however, subsequent studies revealed that $\beta$ is not constant and can vary depending on the physical conditions of the binary system \citep{coriat2009}. For an irradiated disk, $\beta$ ranges from 0.25 in the Rayleigh-Jeans tail to 1.0 in the overlapping or flat region of the spectrum. Conversely, for a viscous disk, the spectrum appears bluer at shorter wavelengths, resulting in $\beta$ values ranging from 0.25 in the Rayleigh-Jeans tail to 0.66 in the overlapping or flat region. Furthermore, \citet{shabhaz2015} demonstrate that $\beta$ can vary between 0.3 and 0.7, depending on the wavelength. This variability in $\beta$ underscores the complexity of the optical/X-ray correlation and the need to account for the diverse physical conditions across different systems.

Based on results from XRB research, we can predict the expected behavior of $\beta$ in AGN by comparing the physical conditions of their emitting regions. In XRBs, the $V$-band typically samples the spectral curvature near the onset of the Rayleigh-Jeans tail. This regime corresponds to effective temperatures of $T_{\mathrm{eff}} \approx 7000$--$20{,}000$ K and yields an observed slope of $\beta \approx 0.6$ \citep{RN6}, consistent with theoretical predictions \citep{RN5}.

For AGN, determining the specific spectral region sampled by the $V$-band is crucial. Given that the accretion disk's spectral peak (the ``big blue bump'') lies in the UV, the $V$-band likely samples the cooler transition region near the onset of the Rayleigh-Jeans tail. This interpretation aligns with standard thin-disk theory and reverberation mapping studies, which characterize the optical emitting region with effective temperatures of $T_{\mathrm{eff}} \sim 10^4$ K \citep{fausnaugh2016}---a thermal regime similar to that observed in XRBs. Consequently, the fact that the AGN sample is well described by the same $L_V \propto a \sqrt{L_X}$ scaling adopted for XRBs (i.e., $\beta = 0.5$) is consistent with this physical similarity. It suggests that despite the vast difference in mass, the optical bands in both populations sample equivalent disk regions, supporting their unified treatment.

Interestingly, despite the flattened slopes observed in individual sources like NGC 2617 and NGC 4151 ($\beta \approx 0.1$; Section~\ref{subsec:edd}), the combined XRB and AGN sample is well described by a single relation with $\beta = 0.5$. This broad agreement is consistent with the theoretical prediction that optical variability in both populations is driven by X-ray reprocessing. The existence of this unified correlation supports our hypothesis that the $V$-band in AGN samples the same thermal reprocessing regime as in XRBs.

However, the observed scatter around this global trend points to significant physical nuances in individual systems. In the following section, we explore the potential drivers of these deviations, specifically examining complexities in the X-ray emission geometry, intrinsic optical contributions from the disk, and the limitations of the standard thin-disk model.

\subsection{Caveats and Contributions to Observed Deviations}

The observed deviations in the optical/X-ray correlation---amounting to a residual scatter of $\sim 0.20$ dex---may arise from variations in the primary observables, as well as from the physical assumptions used to model the system. Identifying the dominant drivers of this scatter is essential for accurately interpreting the deviations and isolating genuine physical differences in accretion geometries from mere observational noise.

\subsubsection{Uncertainties in Primary Observables}

The scatter about the relation has two possible origins: uncertainties in the measured quantities, and genuine source-to-source differences not captured by the simple reprocessing model (e.g., the slope may differ between sources depending on the disk temperature at the reprocessing site). The lag uncertainties alone do not account for the full scatter of 0.201\,dex; an intrinsic component of 0.143\,dex remains (Section~\ref{subsec:global}), indicating that both contributions are present. Uncertainties in the measured time lags ($\tau$) and source distances are among the most significant observational contributors to the scatter. Time lags, which directly determine the characteristic reprocessing radius ($a_R$), carry fractional uncertainties of $\sim$6\%--60\% in our sample (median $\sim$40\%), strongly dependent on the cadence, duration, and signal-to-noise ratio of the monitoring light curves. Similarly, distance measurements are required to convert observed fluxes into intrinsic luminosities ($L_X$ and $L_V$). Distance uncertainties are significant for both populations; for example, the distances to GX~339$-$4 and NGC~3516 carry uncertainties of $\sim$40\% and $\sim$80\%, respectively (Tables~\ref{tab:xrbs} and \ref{tab:agns}). Because $L_V$ depends on the square of the distance whereas the reprocessing term $a_R L_X^{0.5}$ depends only linearly on it, distance errors do not merely slide sources along the correlation line; instead, they shift data points along a steeper direction than the relation, contributing to the residual scatter.

\subsubsection{Methodological and Systematic Effects}

Beyond the primary observables, several methodological assumptions and data-processing steps introduce further scatter. These secondary sources of error include:

\begin{itemize}
    \item Extrapolation of the X-ray Spectrum: To evaluate the correlation accurately, we utilize the broad-band bolometric X-ray luminosity \citep[see][for a discussion]{bernardini2016}. For sources where bolometric values are unavailable, we extrapolate the luminosity from a narrower observed energy band using a power-law photon index ($\Gamma$). Unmodeled spectral complexities, such as high-energy cutoffs, soft X-ray excesses, or variable neutral absorption column densities, can cause this extrapolation to systematically over- or underestimate the true irradiating X-ray flux.

    \item Errors in Host Galaxy Subtraction: For AGN, the observed $V$-band emission is inherently contaminated by the host galaxy's starlight. Although host contributions have been systematically removed, the subtraction process relies on modeling galactic surface brightness profiles. Uncertainties in these profiles can lead to either residual contamination (artificially inflating $L_V$) or over-subtraction (artificially suppressing $L_V$), contributing to vertical scatter on the correlation plot.

\end{itemize}

While these observational and methodological uncertainties account for a portion of the $\sim 0.20$ dex scatter, they cannot fully explain the most extreme outliers. Therefore, large deviations from the correlation line are likely driven by intrinsic, physical differences in the accretion flow, necessitating a closer look at the specific geometries of individual sources.

\subsubsection{Complexities in X-ray Emission Regions in XRBs and AGN}

The location and structure of the corona relative to the accretion disk play a critical role in shaping the observed optical/X-ray correlation. An important caveat when interpreting this correlation in hard-state XRBs is that the observed optical emission is a composite of two components: thermal reprocessing of X-rays in the outer accretion disk and non-thermal synchrotron emission from the relativistic jet \citep{RN6} or from the inner hot accretion flow \citep{veledina2011, veledina2013}. While it is difficult to cleanly isolate the non-thermal contribution to the optical band, both mechanisms scale predictably with X-ray luminosity. Optical reprocessing scales as $L_\mathrm{opt,rep} \propto L^{0.5}_X$, whereas the jet contribution scales as $L_\mathrm{opt,jet} \propto L^{0.7}_X$ \citep{RN6, coriat2009}. Because disk reprocessing is the dominant optical emission source in these systems, with a smaller jet contribution \citep{RN6}, and because the two scalings are nearly indistinguishable in the observed slope, this additional flux does not significantly alter the slope of the correlation. Consequently, the global scaling relation linking XRBs up to AGN remains robust.

However, individual systems can deviate from this baseline correlation, and GRO J1655$-$40 is the clearest XRB example (Figures \ref{fig:correlation} and \ref{fig:inclination}). While its data points may visually appear to deviate toward the right of the main trend, this offset corresponds to a lower optical luminosity for a given X-ray luminosity, by $\sim$0.7 dex. GRO J1655$-$40 is a radio-faint XRB with a weaker jet than typical sources, so its optical emission lacks a non-thermal jet component \citep{RN68}. However, because the jet contributes only a small fraction of the optical light in other hard-state XRBs, its absence can account for only part of this deficit. Known observational uncertainties are also too small to explain it: the extinction ($A_V = 3.7 \pm 0.3$) and distance uncertainties each shift the source by $\lesssim$0.1 dex, and adopting the shorter lag of 14.6\,s \citep{hynes1998} would reduce the offset by only $\sim$0.1 dex. A plausible explanation is that the lag we adopt \citep{obrien2002} was measured during an earlier outburst than our 2005 hard-state fluxes. If the reprocessing geometry differed between the two epochs, the adopted $a_R$ need not represent the reprocessing radius during our observations; matching the relation would require a lag of $\sim$3--4\,s in 2005. Simultaneous lag and flux measurements from a single outburst would test this.

The X-ray emission regions in AGN likely differ from those in stellar-mass counterparts due to their vastly larger physical scales and distinct coronal geometries and processes. Unlike the relatively compact, spherical, or lamppost-like coronae often assumed in XRB models, AGN coronae may exhibit more complex configurations, such as extended, disk-like, or patchy structures (see Section \ref{sec:alt}), influenced by differences in accretion rates, magnetic field geometries, or interactions with surrounding material \citep{haardt1991, kim2024}. These geometries can alter the anisotropy of X-ray emission, affecting its interaction with the accretion disk and the observed optical/X-ray correlation.

Furthermore, the corona in AGN may be more extended or stratified compared to XRBs, with different regions contributing to X-ray production at varying distances from the black hole \citep{wilkins2012}. This extended geometry can lead to uneven illumination of the accretion disk, modifying the efficiency of X-ray reprocessing and potentially influencing the observed correlations. Temporal changes in the coronal structure---such as expansion, contraction, or variations in optical depth---may further modulate these correlations, while the physical mechanisms driving coronal heating and particle acceleration---such as magnetic reconnection, turbulent plasma flows, or shocks induced by outflows and winds---add further complexity to AGN behavior \citep{dimatteo1998, beloborodov1999}.

\subsubsection{Intrinsic Optical Emission from the Accretion Disk and Other Components}

For AGN, optical emission arises not only from X-ray reprocessing but also from intrinsic viscous dissipation within the accretion disk \citep{ss73}. A key distinction is that this viscously heated emission typically varies with a significantly lower amplitude than the reprocessing signals driven by the variable X-ray flux. Consequently, the viscous component effectively adds a quasi-constant baseline to the optical emission. This dilution of the variability signal tends to flatten the observed correlation, which may explain why sources like NGC 2617 and NGC 4151 exhibit flatter slopes with nearly constant optical luminosities. Furthermore, this additional flux shifts data points above the purely reprocessing-driven correlation, making them appear optically brighter. This offset is small compared with the nearly ten-dex range of the correlation, but comparable to its scatter: for example, if the viscous disk contributes $\sim 50\%$ of the total optical emission, the observed flux is only a factor of two ($0.3$ dex) higher than the reprocessing component.

Beyond the accretion disk, additional optical flux may arise from components such as the BLR and the dusty torus. The BLR, composed of dense gas clouds at larger radii surrounding the disk, is photoionized by central X-ray and UV emission, producing line and continuum radiation in the optical and UV bands \citep{netzer2015}. Similarly, the torus, a reservoir of dust at even larger radii, absorbs high-energy radiation and re-emits it primarily in the infrared, with some contribution to optical wavelengths \citep{netzer2015}. Variability in these components, driven by changes in accretion rate or irradiation, can introduce time-delayed optical flux, further complicating the optical/X-ray correlation.

Additionally, the intrinsic viscously heated disk can also vary, correlating with X-ray emission on much longer timescales. In this scenario, as matter propagates through the disk from the outer to the inner regions, optical fluctuations can occur weeks to months before the corresponding X-rays \citep{Lyubarskii1997, Arevalo2006, arevalo2008, breedt2010}. Although such long, negative lags (where the optical leads the X-rays) were easily discarded from our lag analysis, the presence of this varying component nonetheless contributes additional scatter, further complicating the broader optical/X-ray correlation.

\subsubsection{Caveats in Using the Lamppost Model for Time Lag Analysis}

In this analysis, we primarily rely on observed time lags, operating under the fundamental assumption that these lags correspond to the direct light-travel time from the X-ray emitting region to the optical reprocessing radius in the accretion disk. While the global empirical scaling relations do not strictly invoke a specific X-ray coronal geometry, comparing these observed lags to theoretical predictions typically assumes the standard lamppost model---the simplest reprocessing framework. This model posits a compact, isotropic corona positioned directly above the black hole on the rotational axis. Under this assumption, X-ray radiation from a central point-source illuminates the accretion disk, allowing the size of the optical emitting region to be derived directly from continuum reverberation mapping.

However, the technique of translating these time lags directly into physical distances relies heavily on the simplifying assumption that the irradiating flux originates from a single, compact central source \citep[see][]{cackett2021}. In reality, this assumption may break down if the X-ray emitting region is extended, asymmetric, or originates from the inner accretion disk itself. For instance, if the inner disk is geometrically puffed up, or if the outer disk possesses a concave, flared profile, the outer regions can still be effectively irradiated without a suspended lamppost corona. Furthermore, it is critical to note that the relationship between the ionizing X-ray flux and the resulting reprocessed optical flux is not linear; rather, the reprocessed emission scales as $L_V \propto a \ L^{0.5}_X$, governed by the thermal physics of the irradiated disk. Consequently, if the system possesses a complex corona or a non-standard disk geometry, the observed time lag will reflect a complicated convolution of light paths from various emitting regions. In such cases, the straightforward conversion of time lags ($a_R=c\tau$) to physical disk sizes may misrepresent the true geometry and scale of the system.

Several AGN exhibit time lags that challenge the assumptions of the lamppost and standard geometrically thin, optically thick accretion disk model \citep{ss73}. Notably, NGC 4151, NGC 4593, and NGC 5548 display X-ray/UV lags that are several times larger than predicted, as observed by \textit{Swift} \citep{edelso2017, mchard2018}. For instance, the X-ray/UV lag in NGC 4151, initially estimated at 3.4 days, far exceeds the expected light-crossing time between the X-ray and UV-emitting regions. To explain this anomaly, \citet{edelso2017} suggest a two-stage reprocessing scenario for this source: X-rays are first reprocessed into UV emission within an inner toroidal component, which subsequently irradiates the outer disk to produce the observed optical radiation.

Similarly, AGN such as NGC 3516 and NGC 6814 exhibit optical/X-ray time lags that exceed theoretical lamppost predictions by up to an order of magnitude \citep{noda2016, troyer2016}. Notably, both AGN also fall visibly below our main correlation trend line (Section \ref{sec:result}). These dual discrepancies---both in timing (excessively long lags) and energetics (lower relative optical luminosity)---strongly indicate that the true reprocessing geometry in these specific sources fundamentally breaks the assumptions of direct lamppost illumination. If the system lacks an elevated corona, the observed time lag will reflect a much more complicated convolution of light paths. We explore the specific physical geometries that drive these severe deviations in Section \ref{sec:alt}.

Despite these larger-than-expected time lags, the AGN data exhibit a strong correlation with the $L_V \propto a L_X^{0.5}$ relationship derived from XRBs. However, understanding the nature of these deviations requires both improved observational techniques and refined models of the corona and disk. Future advancements, such as X-ray interferometry, may enable direct resolution of the accretion disk's size in AGN, offering a powerful tool to overcome current limitations and refine our understanding of these systems.

\subsection{Alternative Models to explain discrepancies}
\label{sec:alt}

In this section, we explore alternative models that may explain some of the scatter in the correlation and provide deeper insights into the underlying physics.

\subsubsection{Revised Coronal Models to Explain X-ray/UV Lag Excess in AGN}

The observed X-ray/UV time lags in most AGN are a factor of $\sim$ 2.2 (on average) longer than theoretically predicted values \citep{edelso2019}, indicating deviations from the simple lamppost corona and standard thin disk model. This discrepancy suggests the need for alternative explanations.

To address the longer X-ray/UV time lags observed in sources such as NGC 5548, \citet{gardner2017} propose a modified accretion disk and coronal geometry. In this model, a puffed-up region of the accretion disk, driven by UV-induced atomic processes, increases the photosphere's scale height and produces low-temperature Comptonized emission. This region emits far-ultraviolet (FUV) flux, which illuminates the outer thin disk and contributes to variable emission that aligns more closely with the observed UV and optical light curves. However, the extended lags cannot be fully explained by light travel times from X-ray or FUV reprocessing alone. Instead, they may reflect the time required for the vertical structure of the outer disk to respond to changes in FUV illumination.

A critical distinction from the standard lamppost model is the inclusion of a toroidal extreme-UV (EUV) component, which prevents the X-ray corona from directly illuminating the disk. In this scenario, the corona heats the EUV component, which subsequently thermalizes and illuminates the disk, offering a plausible explanation for the observed lag excess. However, this model is still insufficient to explain the significantly longer X-ray/UV lags observed in AGN such as NGC 3516 and NGC 6814, where the time lags are approximately an order of magnitude larger than theoretical predictions. These extreme discrepancies suggest additional physical mechanisms or a fundamentally different accretion geometry, such as the truncated disk with a hot inner flow proposed by \citet{noda2016}, may be at play.

\citet{noda2016} propose another alternative accretion geometry for certain AGN, particularly low-luminosity systems like NGC 3516 and NGC 6814, which departs from the classical thin disk and lamppost corona model. In this scenario, the geometrically thin, optically thick accretion disk is truncated at a larger radius, leaving the inner region filled with a hot, optically thin, radiatively inefficient accretion flow (RIAF), such as an advection-dominated accretion flow (ADAF) \citep{yuan2014}. This inner hot flow produces hard X-rays through inverse Compton scattering, resulting in a harder X-ray spectrum compared to the standard disk-corona configuration. Meanwhile, the truncated thin disk emits predominantly in the UV and optical bands from its cooler outer regions.

Unlike the lamppost model, where a compact corona illuminates the entire thin disk, this model suggests that the interaction between the truncated disk and inner hot flow is more localized and depends on the truncation radius. This geometry naturally explains the longer optical/X-ray time lags observed in some AGN, as the optical emission originates farther out in the truncated disk, increasing the light travel time. Furthermore, it accounts for the relative optical faintness of these specific AGN. In this scenario, the inner region of the optically thick disk physically evaporates to feed the hot inner flow \citep{noda2016}. Consequently, the intrinsic viscously heated thermal continuum from the inner disk, which contributes mainly in the UV, is removed. Moreover, the remaining truncated outer disk intercepts a much smaller fraction of the central X-ray flux compared to an elevated lamppost corona. This dual effect—the absence of inner disk emission and reduced reprocessing efficiency—results in a fainter optical luminosity for a given X-ray flux. This could be the reason that such sources (i.e., NGC 3516 and NGC 6814) appear as outliers that fall below the main optical/X-ray correlation trend line.

Accounting for the extended light travel time caused by disk truncation can also enhance our estimation of the accretion disk size. The current disk size equation (Eq. \ref{eq:lamppost}) is based on the standard thin disk and lamppost coronal model, which assumes a temperature profile of  $T \propto R^{-3/4}$ \citep{ss73}. This derivation relies on the simplifying approximation that the height of the corona above the black hole is negligible compared to the accretion disk radius. However, incorporating a more sophisticated model that accounts for deviations from these assumptions could significantly improve the predictive accuracy of accretion disk size estimates.

Ultimately, validating these alternative models will require the use of general relativistic magnetohydrodynamic (GRMHD) simulations integrated with light echo modeling. For instance, \citet{wong2024} combined GRMHD simulations with radiative transfer techniques to study light echoes through correlations in the total light curve. Such advanced approaches offer critical insights into the complex physics near SMBHs, shedding light on the interactions between the accretion disk and corona. These methods not only refine our theoretical understanding but also provide a framework for interpreting observations, paving the way for future research.

It is important to distinguish between the two revised geometries discussed above. An increased disk scale height mainly lengthens the lags without changing the energetics: sources whose lags exceed the thin-disk prediction, such as NGC 2617 and NGC 4593, nonetheless lie within 0.2 dex of the global relation. Disk truncation, in contrast, alters both the timing and the energetics. NGC 3516 and NGC 6814, the two AGN with the lowest Eddington ratios in our sample ($L_X/L_{\rm Edd} \lesssim 10^{-2}$), lie $\sim$0.85 dex below the relation, as discussed above. For the former geometry, the discrepancy is primarily a ``timing'' excess rather than an ``energetic'' one. Despite the lags being systematically larger than the predictions of the \citet{ss73} model, the inter-band correlations remain remarkably tight. This suggests that the underlying reprocessing mechanism is still driven by the same fundamental accretion physics, even if the transfer function is stretched by a more complex environment or a more distant reprocessing site. Consequently, while the absolute disk size estimates must be scaled by a factor of $\sim$2--3 to match the lag data, the relative trend across different AGN luminosities and black hole masses remains consistent with a unified accretion theory.

\subsection{Sources of interest}

Several other sources could not be included in our analysis, but they are worth mentioning for future studies.

First, \cite{jha2021} identified several AGN sources---Mrk 335, Mrk 486, Mrk 817, PG 1617+175, and PG 2130+099--that exhibit good agreement with the standard accretion disk predictions in the UV-optical regions, based on data from the Zwicky Transient Facility (ZTF) survey. These results support the reprocessing of X-ray photons on the accretion disk near the SMBH. However, we were unable to include these sources in our study due to the absence of quasi-simultaneous optical/X-ray flux measurements and reliable estimates of the light-travel time between the X-ray and UV-emitting regions.

Additionally, MCG+08-11-011 aligns well with the reprocessing lag model \citep{fausna2018}, and the galaxy contribution is negligible, making it a strong candidate for analysis. However, this source was excluded due to the lack of X-ray data and the absence of an X-ray/UV time lag, as the study relied solely on UV and optical data.

Simultaneous X-ray and optical flux data, as well as time lags, are available for MR 2251$-$178 \citep{arevalo2008}. They demonstrated that the small lags between optical and X-ray bands suggest their variations are connected through reprocessing of X-rays, possibly by an optically thick accretion disk, where the light travel time to the reprocessor is the main contributor to the time lags. However, large errors in their lag measurements, such as 1.3 $\pm$ 3.3 days, made them unsuitable for our analysis.

Lastly, Mrk 509 \citep{edelso2019} presents an interesting case where the X-ray flux peaks after the optical emission, contrary to the expectations of the reprocessing model. This behavior violates the assumptions of X-ray-driven reprocessing, preventing the use of this source in our analysis. Investigating why this occurs remains an open question for future studies.

\subsection{Implications of This Study for Modeling AGN Accretion Disk Sizes}

Despite these complexities, the observation that the theoretical correlation for XRBs also holds for AGN implies that the underlying reprocessing mechanisms might be the same in both systems. This correlation provides valuable insights into the geometry of the corona and the structure of accretion disks, offering a powerful tool for studying black hole accretion across mass scales. The X-rays illuminating the disk depend on the location and properties of the corona or hot flow, which generates the hard X-ray emission. If the observed global correlation holds consistently, it could potentially be used to measure not only the size and structure of AGN accretion disks but also the physical characteristics of the corona surrounding SMBHs.

As research on the optical/X-ray correlation matures, reducing scatter in the observed data could enable precise estimates of AGN accretion disk sizes.

By focusing on reprocessing-dominated regimes, the correlation could become a useful diagnostic tool, potentially helping to constrain parameters such as the mass accretion rate and the physical extent of the X-ray corona. This approach would complement established techniques such as reverberation mapping while offering new insights into the coupling between X-ray emission, optical flux, and disk geometry.

Rewriting the correlation equation (Eq. \ref{eq:5}) provides a direct means of estimating the size of the optical-emitting region:

\begin{equation}
    \log a_R = 1.047 \log L_V - 0.500 \log L_X - 6.77,
\end{equation}
where $a_R$ is in cm. Applied to our sample, this relation recovers the measured $\log a_R$ with a root-mean-square residual of 0.21 dex, showing that while the model captures the overall trend, individual estimates scatter by $\sim$0.2 dex.

The intrinsic scatter of Equation~\ref{eq:5} ($\sim$0.143 dex) is competitive with several robust and widely used empirical scaling relations in high-energy astrophysics. For context, the Fundamental Plane of black hole activity \citep{merloni2003, Falcke2004}---a universal scaling law connecting X-ray luminosity, radio luminosity, and black hole mass across both XRBs and AGN---exhibits a comparable intrinsic scatter, often ranging from $\sim$ 0.1 to 0.4 dex depending on sample selection and fitting methods \citep{Plotkin2012}. Similarly, the global relation linking the characteristic X-ray variability timescales to black hole mass and accretion rate across both XRBs and AGN shows a comparable level of intrinsic dispersion \citep{mchardy2006}. Therefore, while the residual discrepancies in our correlation likely stem from unmodeled complexities or systematic structural patterns (such as the idiosyncratic reprocessing geometries discussed earlier), the baseline accuracy of Equation \ref{eq:5} already matches the predictive precision of the most established global accretion scaling laws.

Recent studies have also sought to estimate AGN accretion disk sizes using continuum reverberation mapping. For instance, \citet{jha2021} measured time lags to constrain disk dimensions, finding them to be systematically larger than standard theoretical predictions. However, their analysis relied exclusively on lags between three optical bands ($g$, $r$, and $i$), lacking simultaneous coverage in the UV and X-ray regimes. Because the primary heating source that drives thermal reprocessing originates in the central corona, relying solely on intra-optical lags adds complexity when attempting to map the full disk-corona energy transfer. In contrast, by directly incorporating X-ray luminosities and X-ray-to-optical lags in our study, it becomes significantly more straightforward to isolate the reprocessed component. This direct baseline allows for a much cleaner prediction of the reprocessing behavior and provides a clearer view of the global correlation connecting the corona and the outer disk.

\subsection{The Observational ``Sweet Spot'' for Reprocessing Signatures}

By combining our results with established AGN scaling relations, we can make general predictions for the observability of disk reprocessing across different accretion regimes. The intrinsic disk--corona relation for radiatively efficient AGN follows $L_{\mathrm{X}} \propto L_{\mathrm{UV}}^{0.64}$ \citep[e.g.,][]{Lusso2016}, while Equation~\ref{eq:5} gives $L_{\mathrm{V}} \propto L_{\mathrm{X}}^{0.48}$ at fixed $a_R$. Together, these imply an approximate dependence $L_{\mathrm{V,rep}} \propto L_{\mathrm{UV}}^{0.31}$.

These relations originate from distinct accretion regimes: the former reflects intrinsic disk--corona coupling in radiatively efficient quasars, whereas the latter spans a broad range of luminosities down to $< 10^{-4} \, L_\mathrm{Edd}$. Nevertheless, this combined scaling provides a useful heuristic baseline. In stellar-mass XRBs, an un-irradiated, viscously heated disk produces a characteristically shallow optical/X-ray correlation ($L_\mathrm{V} \propto L_\mathrm{X}^{\sim 0.2\text{--}0.4}$) \citep[][]{RN6, Russell2007, Armas2013, Yang2022}. Because X-ray thermal reprocessing predicts a steeper dependence ($L_\mathrm{V} \propto L_\mathrm{X}^{0.5}$), reprocessed light grows more rapidly with X-ray luminosity than intrinsic viscous optical emission, causing reprocessing to dominate the optical output of the disk as the accretion rate increases.

At the high-luminosity end, however, the shallow dependence $L_{\mathrm{V,rep}} \propto L_{\mathrm{UV}}^{0.31}$ means that reprocessed light grows more slowly than the intrinsic disk emission, so observational signatures of reprocessing---such as correlated multi-band variability and inter-band time lags---are expected to be more prominent in lower-luminosity systems. However, detecting these signatures at the lowest luminosities is observationally challenging due to reduced signal-to-noise in both X-ray and optical data.

This leads to the existence of an observational ``sweet spot'': sources must be sufficiently bright to enable high-quality variability measurements, yet not so luminous that the reprocessed signal is overwhelmed by intrinsic disk emission. This may help explain why reprocessing signatures are most readily detected in moderate-luminosity Seyfert 1 galaxies, such as those in our sample.

A key prediction follows from this framework: the amplitude of optical variability driven by X-ray reprocessing, as well as the strength of X-ray/optical correlations, should systematically decrease with increasing luminosity and Eddington ratio. This trend can be directly tested with future reverberation mapping campaigns spanning a wide range of accretion states.

\section{Conclusion} \label{sec:con}

In this study, we investigated whether the $L_V$–$a L_X^{0.5}$ correlation, observed in XRBs, can be extended to AGN. We collected a comprehensive multi-wavelength dataset of X-ray and optical fluxes from five XRBs and seven AGN, incorporating time lag measurements to characterize the relationship. All X-ray fluxes were converted to bolometric luminosities, and the time lags between two bands were used to infer the reprocessing radius on the accretion disk. A strong correlation between $L_V$–$a L_X^{0.5}$ was identified for both XRBs and AGN, spanning almost ten orders of magnitude in $L_V$ and encompassing black hole systems of vastly different sizes, with a total scatter of $0.20$ dex ($0.18$ dex when corrected for inclination), corresponding to an intrinsic scatter of $0.143$ dex ($0.127$ dex) once the lag uncertainties are accounted for.

The robustness of this scaling is evident when comparing the raw and inclination-corrected data. Without inclination corrections, the sample exhibits a correlation coefficient of $r \approx$ 0.9979 (or scatter $\sigma \approx$ 0.20). When correcting for system inclination, the correlation tightens slightly ($r \approx$ 0.9983 with scatter of $\sigma \approx$ 0.18), consistent with the adopted $\beta = 0.5$ scaling. This agreement suggests that AGN share a similar X-ray emission geometry with XRBs, where the optical emitting region is governed by X-ray reprocessing rather than intrinsic disk fluctuations alone. To further test this hypothesis, additional validation with XRB data is necessary. Comparing observed time lags with independently measurable accretion disk sizes in XRBs can provide insight into the specific radii of the disk responsible for optical emission.

These findings have broad implications for the unification of black hole accretion physics. Our results indicate that the same reprocessing mechanism can account for disk irradiation in both stellar-mass and supermassive black holes. While direct measurement of accretion disk sizes in AGN remains challenging, the time lag--size relation examined here offers a viable method for estimating these parameters by scaling from well-constrained XRB systems. Ultimately, this work provides evidence that accretion physics scales predictably with mass. By establishing a unified scaling for the optical continuum, it adds a crucial geometric dimension to established unification frameworks like the Fundamental Plane \citep{merloni2003} and the X-ray variability plane \citep{mchardy2006}, complementing previous adaptations that utilized optical emission lines \citep{saikia2015}.

\begin{acknowledgments}
This material is based upon work supported by Tamkeen under the NYUAD Research Institute grant CG014. We thank Rob Fender for useful discussions regarding global correlations and disk size estimates. We also thank Dan Bramich for insightful discussions and helpful comments.
\end{acknowledgments}

\vspace{5mm}
\facilities{CTIO:1.3m, INTEGRAL (IBIS), LCOGT, Liverpool:2m, NICER, NTT, NuSTAR, RXTE (ASM, PCA), Suzaku, Swift (XRT, UVOT)}

\bibliographystyle{aasjournal}
\bibliography{sample631}

\begin{thebibliography}{}
\expandafter\ifx\csname natexlab\endcsname\relax\def\natexlab#1{#1}\fi
\providecommand{\url}[1]{\href{#1}{#1}}
\providecommand{\dodoi}[1]{doi:~\href{http://doi.org/#1}{\nolinkurl{#1}}}
\providecommand{\doeprint}[1]{\href{http://ascl.net/#1}{\nolinkurl{http://ascl.net/#1}}}
\providecommand{\doarXiv}[1]{\href{https://arxiv.org/abs/#1}{\nolinkurl{https://arxiv.org/abs/#1}}}

\bibitem[{{Aftab} \& {Paul}(2024)}]{2024NewA..10602109A}
{Aftab}, N., \& {Paul}, B. 2024, \na, 106, 102109,
  \dodoi{10.1016/j.newast.2023.102109}

\bibitem[{{Ahmad} {et~al.}(2018){Ahmad}, {Misra}, {Iqbal}, {Maqbool}, \&
  {Hamid}}]{2018NewA...58...84A}
{Ahmad}, N., {Misra}, R., {Iqbal}, N., {Maqbool}, B., \& {Hamid}, M. 2018, \na,
  58, 84, \dodoi{10.1016/j.newast.2017.07.011}

\bibitem[{{Arcodia} {et~al.}(2019){Arcodia}, {Merloni}, {Nandra}, \&
  {Ponti}}]{Arcodia2019}
{Arcodia}, R., {Merloni}, A., {Nandra}, K., \& {Ponti}, G. 2019, \aap, 628,
  A135, \dodoi{10.1051/0004-6361/201935874}

\bibitem[{{Arcodia} {et~al.}(2020){Arcodia}, {Ponti}, {Merloni}, \&
  {Nandra}}]{Arcodia2020}
{Arcodia}, R., {Ponti}, G., {Merloni}, A., \& {Nandra}, K. 2020, \aap, 638,
  A100, \dodoi{10.1051/0004-6361/202037969}

\bibitem[{{Ar{\'e}valo} \& {Uttley}(2006)}]{Arevalo2006}
{Ar{\'e}valo}, P., \& {Uttley}, P. 2006, \mnras, 367, 801,
  \dodoi{10.1111/j.1365-2966.2006.09989.x}

\bibitem[{{Armas Padilla} {et~al.}(2013){Armas Padilla}, {Degenaar}, {Russell},
  \& {Wijnands}}]{Armas2013}
{Armas Padilla}, M., {Degenaar}, N., {Russell}, D.~M., \& {Wijnands}, R. 2013,
  \mnras, 428, 3083, \dodoi{10.1093/mnras/sts255}

\bibitem[{Arévalo {et~al.}(2008)Arévalo, Uttley, Kaspi, Breedt, Lira, \&
  McHardy}]{arevalo2008}
Arévalo, P., Uttley, P., Kaspi, S., {et~al.} 2008, Monthly Notices of the
  Royal Astronomical Society, 389, 1479,
  \dodoi{10.1111/j.1365-2966.2008.13719.x}

\bibitem[{{Atri} {et~al.}(2020){Atri}, {Miller-Jones}, {Bahramian}, {Plotkin},
  {Deller}, {Jonker}, {Maccarone}, {Sivakoff}, {Soria}, {Altamirano},
  {Belloni}, {Fender}, {Koerding}, {Maitra}, {Markoff}, {Migliari}, {Russell},
  {Russell}, {Sarazin}, {Tetarenko}, \& {Tudose}}]{2020MNRAS.493L..81A}
{Atri}, P., {Miller-Jones}, J.~C.~A., {Bahramian}, A., {et~al.} 2020, \mnras,
  493, L81, \dodoi{10.1093/mnrasl/slaa010}

\bibitem[{Baglio {et~al.}(2018)Baglio, Russell, Casella, Al~Noori, Al~Yazeedi,
  Belloni, Buckley, Cadolle~Bel, Ceccobello, Corbel, Coti~Zelati, Díaz~Trigo,
  Fender, Gallo, Gandhi, Homan, Koljonen, Lewis, Maccarone, Malzac, Markoff,
  Miller-Jones, O’Brien, Russell, Saikia, Shahbaz, Sivakoff, Soria, Testa,
  Tetarenko, van~den Ancker, \& Vincentelli}]{baglio2018}
Baglio, C.~M., Russell, D.~M., Casella, P., {et~al.} 2018, The Astrophysical
  Journal, 867, 114, \dodoi{10.3847/1538-4357/aae532}

\bibitem[{Bahcall {et~al.}(1972)Bahcall, Kozlovsky, \& Salpeter}]{bahcall1972}
Bahcall, J.~N., Kozlovsky, B.-Z., \& Salpeter, E.~E. 1972, The Astrophysical
  Journal, 171, 467, \dodoi{10.1086/151300}

\bibitem[{{Barnier} {et~al.}(2023){Barnier}, {Petrucci}, {Ferreira}, \&
  {Marcel}}]{Barnier2023}
{Barnier}, S., {Petrucci}, P.-O., {Ferreira}, J., \& {Marcel}, G. 2023,
  Astronomische Nachrichten, 344, e20230020, \dodoi{10.1002/asna.20230020}

\bibitem[{{Barthelmy} {et~al.}(2005){Barthelmy}, {Barbier}, {Cummings},
  {Fenimore}, {Gehrels}, {Hullinger}, {Krimm}, {Markwardt}, {Palmer},
  {Parsons}, {Sato}, {Suzuki}, {Takahashi}, {Tashiro}, \&
  {Tueller}}]{Barthelmy2005}
{Barthelmy}, S.~D., {Barbier}, L.~M., {Cummings}, J.~R., {et~al.} 2005, \ssr,
  120, 143, \dodoi{10.1007/s11214-005-5096-3}

\bibitem[{{Beloborodov}(1999)}]{beloborodov1999}
{Beloborodov}, A.~M. 1999, \apjl, 510, L123, \dodoi{10.1086/311810}

\bibitem[{Bentz {et~al.}(2019)Bentz, Ferrarese, Onken, Peterson, \&
  Valluri}]{bentz2019}
Bentz, M.~C., Ferrarese, L., Onken, C.~A., Peterson, B.~M., \& Valluri, M.
  2019, The Astrophysical Journal, 885, 161, \dodoi{10.3847/1538-4357/ab48fb}

\bibitem[{Bentz {et~al.}(2009)Bentz, Walsh, Barth, Baliber, Bennert, Canalizo,
  Filippenko, Ganeshalingam, Gates, Greene, Hidas, Hiner, Lee, Li, Malkan,
  Minezaki, Sakata, Serduke, Silverman, Steele, Stern, Street, Thornton, Treu,
  Wang, Woo, \& Yoshii}]{bentz2009}
Bentz, M.~C., Walsh, J.~L., Barth, A.~J., {et~al.} 2009, The Astrophysical
  Journal, 705, 199, \dodoi{10.1088/0004-637X/705/1/199}

\bibitem[{Bernardini {et~al.}(2016)Bernardini, Russell, Kolojonen, Stella,
  Hynes, \& Corbel}]{bernardini2016}
Bernardini, F., Russell, D.~M., Kolojonen, K. I.~I., {et~al.} 2016, The
  Astrophysical Journal, 826, 149, \dodoi{10.3847/0004-637X/826/2/149}

\bibitem[{{Bernardini} {et~al.}(2016){Bernardini}, {Russell}, {Shaw}, {Lewis},
  {Charles}, {Koljonen}, {Lasota}, \& {Casares}}]{Bernardini2016b}
{Bernardini}, F., {Russell}, D.~M., {Shaw}, A.~W., {et~al.} 2016, \apjl, 818,
  L5, \dodoi{10.3847/2041-8205/818/1/L5}

\bibitem[{Blandford \& McKee(1982)}]{RN56}
Blandford, R.~D., \& McKee, C.~F. 1982, The Astrophysical Journal, 255, 419,
  \dodoi{10.1086/159843}

\bibitem[{Bontà {et~al.}(2020)Bontà, Peterson, Bentz, Brandt, Ciroi, Rosa,
  Alvarez, Grier, Hall, Hernández~Santisteban, Ho, Homayouni, Horne, Kochanek,
  Li, Morelli, Pizzella, Pogge, Schneider, Shen, Trump, \& Vestergaard}]{RN29}
Bontà, E.~D., Peterson, B.~M., Bentz, M.~C., {et~al.} 2020, The Astrophysical
  Journal, 903, \dodoi{10.3847/1538-4357/abbc1c}

\bibitem[{{Breedt} {et~al.}(2009){Breedt}, {Ar{\'e}valo}, {McHardy}, {Uttley},
  {Sergeev}, {Minezaki}, {Yoshii}, {Gaskell}, {Cackett}, {Horne}, \&
  {Koshida}}]{2009MNRAS.394..427B}
{Breedt}, E., {Ar{\'e}valo}, P., {McHardy}, I.~M., {et~al.} 2009, \mnras, 394,
  427, \dodoi{10.1111/j.1365-2966.2008.14302.x}

\bibitem[{Breedt {et~al.}(2010)Breedt, McHardy, Arévalo, Uttley, Sergeev,
  Minezaki, Yoshii, Sakata, Lira, \& Chesnok}]{breedt2010}
Breedt, E., McHardy, I.~M., Arévalo, P., {et~al.} 2010, Monthly Notices of the
  Royal Astronomical Society, 403, 605,
  \dodoi{10.1111/j.1365-2966.2009.16146.x}

\bibitem[{{Bright} {et~al.}(2025){Bright}, {Fender}, {Russell}, {Motta}, {Man},
  {van den Eijnden}, {Alabarta}, {Crook-Mansour}, {Baglio}, {Green}, {Heywood},
  {Lewis}, {Saikia}, {Scott}, \& {Titterington}}]{Bright2025}
{Bright}, J.~S., {Fender}, R., {Russell}, D.~M., {et~al.} 2025, \mnras, 541,
  1851, \dodoi{10.1093/mnras/staf1098}

\bibitem[{Brocksopp {et~al.}(2010)Brocksopp, Jonker, Maitra, Krimm, Pooley,
  Ramsay, \& Zurita}]{RN66}
Brocksopp, C., Jonker, P.~G., Maitra, D., {et~al.} 2010, Monthly Notices of the
  Royal Astronomical Society, 404, 908,
  \dodoi{10.1111/j.1365-2966.2010.16323.x}

\bibitem[{Brok {et~al.}(2015)Brok, Seth, Barth, Carson, Neumayer, Cappellari,
  Debattista, Ho, Hood, \& McDermid}]{brok2015}
Brok, M.~d., Seth, A.~C., Barth, A.~J., {et~al.} 2015, The Astrophysical
  Journal, 809, 101, \dodoi{10.1088/0004-637X/809/1/101}

\bibitem[{Buxton {et~al.}(2012)Buxton, Bailyn, Capelo, Chatterjee, Dinçer,
  Kalemci, \& Tomsick}]{RN65}
Buxton, M.~M., Bailyn, C.~D., Capelo, H.~L., {et~al.} 2012, The Astronomical
  Journal, 143, \dodoi{10.1088/0004-6256/143/6/130}

\bibitem[{Cackett {et~al.}(2021)Cackett, Bentz, \& Kara}]{cackett2021}
Cackett, E.~M., Bentz, M.~C., \& Kara, E. 2021, iScience, 24, 102557,
  \dodoi{10.1016/j.isci.2021.102557}

\bibitem[{{Cackett} {et~al.}(2018){Cackett}, {Chiang}, {McHardy}, {Edelson},
  {Goad}, {Horne}, \& {Korista}}]{cackett2018}
{Cackett}, E.~M., {Chiang}, C.-Y., {McHardy}, I., {et~al.} 2018, \apj, 857, 53,
  \dodoi{10.3847/1538-4357/aab4f7}

\bibitem[{{Cackett} {et~al.}(2007){Cackett}, {Horne}, \&
  {Winkler}}]{2007MNRAS.380..669C}
{Cackett}, E.~M., {Horne}, K., \& {Winkler}, H. 2007, \mnras, 380, 669,
  \dodoi{10.1111/j.1365-2966.2007.12098.x}

\bibitem[{{Cackett} {et~al.}(2020){Cackett}, {Gelbord}, {Li}, {Horne}, {Wang},
  {Barth}, {Bai}, {Bian}, {Carroll}, {Du}, {Edelson}, {Goad}, {Ho}, {Hu},
  {Khatu}, {Luo}, {Miller}, \& {Yuan}}]{cackett2020}
{Cackett}, E.~M., {Gelbord}, J., {Li}, Y.-R., {et~al.} 2020, \apj, 896, 1,
  \dodoi{10.3847/1538-4357/ab91b5}

\bibitem[{Cadolle~Bel {et~al.}(2007)Cadolle~Bel, Ribó, Rodriguez, Chaty,
  Corbel, Goldwurm, Frontera, Farinelli, D’Avanzo, Tarana, Ubertini, Laurent,
  Goldoni, \& Mirabel}]{cadoll2007}
Cadolle~Bel, M., Ribó, M., Rodriguez, J., {et~al.} 2007, The Astrophysical
  Journal, 659, 549, \dodoi{10.1086/512004}

\bibitem[{{Callanan} {et~al.}(1995){Callanan}, {Garcia}, {McClintock}, {Zhao},
  {Remillard}, {Bailyn}, {Orosz}, {Harmon}, \&
  {Paciesas}}]{1995ApJ...441..786C}
{Callanan}, P.~J., {Garcia}, M.~R., {McClintock}, J.~E., {et~al.} 1995, \apj,
  441, 786, \dodoi{10.1086/175402}

\bibitem[{Cameron {et~al.}(2012)Cameron, McHardy, Dwelly, Breedt, Uttley, Lira,
  \& Arevalo}]{cameron2012}
Cameron, D.~T., McHardy, I., Dwelly, T., {et~al.} 2012, Monthly Notices of the
  Royal Astronomical Society, 422, 902,
  \dodoi{10.1111/j.1365-2966.2012.20677.x}

\bibitem[{Cardelli {et~al.}(1989)Cardelli, Clayton, \& Mathis}]{cardel1989}
Cardelli, J.~A., Clayton, G.~C., \& Mathis, J.~S. 1989, The Astrophysical
  Journal, 345, 245, \dodoi{10.1086/167900}

\bibitem[{{Casella} {et~al.}(2010){Casella}, {Maccarone}, {O'Brien}, {Fender},
  {Russell}, {van der Klis}, {Pe'Er}, {Maitra}, {Altamirano}, {Belloni},
  {Kanbach}, {Klein-Wolt}, {Mason}, {Soleri}, {Stefanescu}, {Wiersema}, \&
  {Wijnands}}]{casella2010}
{Casella}, P., {Maccarone}, T.~J., {O'Brien}, K., {et~al.} 2010, \mnras, 404,
  L21, \dodoi{10.1111/j.1745-3933.2010.00826.x}

\bibitem[{{Chaty} {et~al.}(2003){Chaty}, {Haswell}, {Malzac}, {Hynes},
  {Shrader}, \& {Cui}}]{2003MNRAS.346..689C}
{Chaty}, S., {Haswell}, C.~A., {Malzac}, J., {et~al.} 2003, \mnras, 346, 689,
  \dodoi{10.1111/j.1365-2966.2003.07115.x}

\bibitem[{Chen {et~al.}(1997)Chen, Shrader, \& Livio}]{chen1997}
Chen, W., Shrader, C.~R., \& Livio, M. 1997, The Astrophysical Journal, 491,
  312, \dodoi{10.1086/304921}

\bibitem[{{Chira} {et~al.}(2026){Chira}, {Georgakakis}, {Ruiz}, {Chen},
  {Buchner}, {Rankine}, {Kammoun}, {Aydar}, {Salvato}, {Merloni}, \&
  {Krumpe}}]{Chira2026}
{Chira}, M., {Georgakakis}, A., {Ruiz}, A., {et~al.} 2026, \mnras, 545,
  staf1905, \dodoi{10.1093/mnras/staf1905}

\bibitem[{Clavel {et~al.}(1983)Clavel, Joly, Collin-Souffrin, Bergeron, \&
  Penston}]{clabel1983}
Clavel, J., Joly, M., Collin-Souffrin, S., Bergeron, J., \& Penston, M.~V.
  1983, Monthly Notices of the Royal Astronomical Society, 202, 85,
  \dodoi{10.1093/mnras/202.1.85}

\bibitem[{Corbel \& Fender(2002)}]{corbel2002}
Corbel, S., \& Fender, R.~P. 2002, The Astrophysical Journal, 573, L35,
  \dodoi{10.1086/341870}

\bibitem[{Coriat {et~al.}(2009)Coriat, Corbel, Buxton, Bailyn, Tomsick,
  Körding, \& Kalemci}]{coriat2009}
Coriat, M., Corbel, S., Buxton, M.~M., {et~al.} 2009, Monthly Notices of the
  Royal Astronomical Society, 400, 123,
  \dodoi{10.1111/j.1365-2966.2009.15461.x}

\bibitem[{{Cunningham}(1976)}]{1976ApJ...208..534C}
{Cunningham}, C. 1976, \apj, 208, 534, \dodoi{10.1086/154636}

\bibitem[{De~Rosa {et~al.}(2015)De~Rosa, Peterson, Ely, Kriss, Crenshaw, Horne,
  Korista, Netzer, Pogge, Arévalo, Barth, Bentz, Brandt, Breeveld, Brewer,
  Dalla~Bontà, De~Lorenzo-Cáceres, Denney, Dietrich, Edelson, Evans,
  Fausnaugh, Gehrels, Gelbord, Goad, Grier, Grupe, Hall, Kaastra, Kelly,
  Kennea, Kochanek, Lira, Mathur, McHardy, Nousek, Pancoast, Papadakis, Pei,
  Schimoia, Siegel, Starkey, Treu, Uttley, Vaughan, Vestergaard, Villforth,
  Yan, Young, \& Zu}]{RN51}
De~Rosa, G., Peterson, B.~M., Ely, J., {et~al.} 2015, The Astrophysical
  Journal, 806, 128, \dodoi{10.1088/0004-637x/806/1/128}

\bibitem[{Denney {et~al.}(2010)Denney, Peterson, Pogge, Adair, Atlee, Au-Yong,
  Bentz, Bird, Brokofsky, Chisholm, Comins, Dietrich, Doroshenko, Eastman,
  Efimov, Ewald, Ferbey, Gaskell, Hedrick, Jackson, Klimanov, Klimek, Kruse,
  Ladéroute, Lamb, Leighly, Minezaki, Nazarov, Onken, Petersen, Peterson,
  Poindexter, Sakata, Schlesinger, Sergeev, Skolski, Stieglitz, Tobin,
  Unterborn, Vestergaard, Watkins, Watson, \& Yoshii}]{denney2010}
Denney, K.~D., Peterson, B.~M., Pogge, R.~W., {et~al.} 2010, The Astrophysical
  Journal, 721, 715, \dodoi{10.1088/0004-637x/721/1/715}

\bibitem[{di~Matteo(1998)}]{dimatteo1998}
di~Matteo, T. 1998, Monthly Notices of the Royal Astronomical Society, 299,
  L15, \dodoi{10.1046/j.1365-8711.1998.01950.x}

\bibitem[{Dickey \& Lockman(1990)}]{dickey1990}
Dickey, J.~M., \& Lockman, F.~J. 1990, Annual Review of Astronomy and
  Astrophysics, 28, 215, \dodoi{10.1146/annurev.aa.28.090190.001243}

\bibitem[{Dinçer {et~al.}(2014)Dinçer, Kalemci, Tomsick, Buxton, \&
  Bailyn}]{dincer2014}
Dinçer, T., Kalemci, E., Tomsick, J.~A., Buxton, M.~M., \& Bailyn, C.~D. 2014,
  The Astrophysical Journal, 795, 74, \dodoi{10.1088/0004-637X/795/1/74}

\bibitem[{Dunn {et~al.}(2010)Dunn, Fender, Körding, Belloni, \&
  Cabanac}]{dunn2010}
Dunn, R. J.~H., Fender, R.~P., Körding, E.~G., Belloni, T., \& Cabanac, C.
  2010, Monthly Notices of the Royal Astronomical Society, 403, 61,
  \dodoi{10.1111/j.1365-2966.2010.16114.x}

\bibitem[{Edelson {et~al.}(2015)Edelson, Gelbord, Horne, McHardy, Peterson,
  Arévalo, Breeveld, Rosa, Evans, Goad, Kriss, Brandt, Gehrels, Grupe, Kennea,
  Kochanek, Nousek, Papadakis, Siegel, Starkey, Uttley, Vaughan, Young, Barth,
  Bentz, Brewer, Crenshaw, Dalla~Bontà, Cáceres, Denney, Dietrich, Ely,
  Fausnaugh, Grier, Hall, Kaastra, Kelly, Korista, Lira, Mathur, Netzer,
  Pancoast, Pei, Pogge, Schimoia, Treu, Vestergaard, Villforth, Yan, \&
  Zu}]{RN41}
Edelson, R., Gelbord, J.~M., Horne, K., {et~al.} 2015, The Astrophysical
  Journal, 806, \dodoi{10.1088/0004-637x/806/1/129}

\bibitem[{Edelson {et~al.}(2017)Edelson, Gelbord, Cackett, Connolly, Done,
  Fausnaugh, Gardner, Gehrels, Goad, Horne, McHardy, Peterson, Vaughan,
  Vestergaard, Breeveld, Barth, Bentz, Bottorff, Brandt, Crawford, Bontà,
  Emmanoulopoulos, Evans, Jaimes, Filippenko, Ferland, Grupe, Joner, Kennea,
  Korista, Krimm, Kriss, Leonard, Mathur, Netzer, Nousek, Page,
  Romero-Colmenero, Siegel, Starkey, Treu, Vogler, Winkler, \&
  Zheng}]{edelso2017}
Edelson, R., Gelbord, J., Cackett, E., {et~al.} 2017, The Astrophysical
  Journal, 840, 41, \dodoi{10.3847/1538-4357/aa6890}

\bibitem[{Edelson {et~al.}(2019)Edelson, Gelbord, Cackett, Peterson, Horne,
  Barth, Starkey, Bentz, Brandt, Goad, Joner, Korista, Netzer, Page, Uttley,
  Vaughan, Breeveld, Cenko, Done, Evans, Fausnaugh, Ferland, Gonzalez-Buitrago,
  Gropp, Grupe, Kaastra, Kennea, Kriss, Mathur, Mehdipour, Mudd, Nousek,
  Schmidt, Vestergaard, \& Villforth}]{edelso2019}
---. 2019, Astrophysical Journal, 870, 123, \dodoi{10.3847/1538-4357/aaf3b4}

\bibitem[{{Falcke} {et~al.}(2004){Falcke}, {K{\"o}rding}, \&
  {Markoff}}]{Falcke2004}
{Falcke}, H., {K{\"o}rding}, E., \& {Markoff}, S. 2004, \aap, 414, 895,
  \dodoi{10.1051/0004-6361:20031683}

\bibitem[{Fausnaugh {et~al.}(2016)Fausnaugh, Denney, Barth, Bentz, Bottorff,
  Carini, Croxall, De~Rosa, Goad, Horne, Joner, Kaspi, Kim, Klimanov, Kochanek,
  Leonard, Netzer, Peterson, Schnülle, Sergeev, Vestergaard, Zheng, Zu,
  Anderson, Arévalo, Bazhaw, Borman, Boroson, Brandt, Breeveld, Brewer,
  Cackett, Crenshaw, Dalla~Bonta, De~Lorenzo-Cáceres, Dietrich, Edelson,
  Efimova, Ely, Evans, Filippenko, Flatland, Gehrels, Geier, Gelbord, Gonzalez,
  Gorjian, Grier, Grupe, Hall, Hicks, Horenstein, Hutchison, Im, Jensen, Jones,
  Kaastra, Kelly, Kennea, Kim, Korista, Kriss, Lee, Lira, MacInnis,
  Manne-Nicholas, Mathur, McHardy, Montouri, Musso, Nazarov, Norris, Nousek,
  Okhmat, Pancoast, Papadakis, Parks, Pei, Pogge, Pott, Rafter, Rix, Saylor,
  Schimoia, Siegel, Spencer, Starkey, Sung, Teems, Treu, Turner, Uttley,
  Villforth, Weiss, Woo, Yan, \& Young}]{fausnaugh2016}
Fausnaugh, M.~M., Denney, K.~D., Barth, A.~J., {et~al.} 2016, Astrophysical
  Journal, 821, 56, \dodoi{10.3847/0004-637x/821/1/56}

\bibitem[{Fausnaugh {et~al.}(2018)Fausnaugh, Starkey, Horne, Kochanek,
  Peterson, Bentz, Denney, Grier, Grupe, Pogge, De~Rosa, Adams, Barth, Beatty,
  Bhattacharjee, Borman, Boroson, Bottorff, Brown, Brown, Brotherton, Coker,
  Crawford, Croxall, Eftekharzadeh, Eracleous, Joner, Henderson, Holoien,
  Hutchison, Kaspi, Kim, King, Li, Lochhaas, Ma, MacInnis, Manne-Nicholas,
  Mason, Montuori, Mosquera, Mudd, Musso, Nazarov, Nguyen, Okhmat, Onken,
  Ou-Yang, Pancoast, Pei, Penny, Poleski, Rafter, Romero-Colmenero, Runnoe,
  Sand, Schimoia, Sergeev, Shappee, Simonian, Somers, Spencer, Stevens, Tayar,
  Treu, Valenti, Van~Saders, Villanueva, Villforth, Weiss, Winkler, \&
  Zhu}]{fausna2018}
Fausnaugh, M.~M., Starkey, D.~A., Horne, K., {et~al.} 2018, The Astrophysical
  Journal, 854, 107, \dodoi{10.3847/1538-4357/aaaa2b}

\bibitem[{{Fender} {et~al.}(2004){Fender}, {Belloni}, \& {Gallo}}]{Fender2004}
{Fender}, R.~P., {Belloni}, T.~M., \& {Gallo}, E. 2004, \mnras, 355, 1105,
  \dodoi{10.1111/j.1365-2966.2004.08384.x}

\bibitem[{{Ferreira} {et~al.}(2006){Ferreira}, {Petrucci}, {Henri},
  {Saug{\'e}}, \& {Pelletier}}]{Ferreira2006}
{Ferreira}, J., {Petrucci}, P.-O., {Henri}, G., {Saug{\'e}}, L., \&
  {Pelletier}, G. 2006, \aap, 447, 813, \dodoi{10.1051/0004-6361:20052689}

\bibitem[{Filippenko \& Ho(2003)}]{filippenko2003}
Filippenko, A.~V., \& Ho, L.~C. 2003, The Astrophysical Journal, 588, L13,
  \dodoi{10.1086/375361}

\bibitem[{Filippenko \& Sargent(1989)}]{filippenko1989}
Filippenko, A.~V., \& Sargent, W. L.~W. 1989, The Astrophysical Journal, 342,
  L11, \dodoi{10.1086/185472}

\bibitem[{Fischer {et~al.}(2013)Fischer, Crenshaw, Kraemer, \&
  Schmitt}]{fischer2013}
Fischer, T.~C., Crenshaw, D.~M., Kraemer, S.~B., \& Schmitt, H.~R. 2013, The
  Astrophysical Journal Supplement Series, 209, 1,
  \dodoi{10.1088/0067-0049/209/1/1}

\bibitem[{{Froning} {et~al.}(2014){Froning}, {Maccarone}, {France}, {Winter},
  {Robinson}, {Hynes}, \& {Lewis}}]{fronin2014}
{Froning}, C.~S., {Maccarone}, T.~J., {France}, K., {et~al.} 2014, \apj, 780,
  48, \dodoi{10.1088/0004-637X/780/1/48}

\bibitem[{{Gandhi} {et~al.}(2009){Gandhi}, {Horst}, {Smette}, {H{\"o}nig},
  {Comastri}, {Gilli}, {Vignali}, \& {Duschl}}]{2009A&A...502..457G}
{Gandhi}, P., {Horst}, H., {Smette}, A., {et~al.} 2009, \aap, 502, 457,
  \dodoi{10.1051/0004-6361/200811368}

\bibitem[{Gandhi {et~al.}(2010)Gandhi, Dhillon, Durant, Fabian, Kubota,
  Makishima, Malzac, Marsh, Miller, Shahbaz, Spruit, \& Casella}]{RN232}
Gandhi, P., Dhillon, V.~S., Durant, M., {et~al.} 2010, Monthly Notices of the
  Royal Astronomical Society, 407, 2166,
  \dodoi{10.1111/j.1365-2966.2010.17083.x}

\bibitem[{Gandhi {et~al.}(2011)Gandhi, Blain, Russell, Casella, Malzac, Corbel,
  D’Avanzo, Lewis, Markoff, Cadolle~Bel, Goldoni, Wachter, Khangulyan, \&
  Mainzer}]{RN231}
Gandhi, P., Blain, A.~W., Russell, D.~M., {et~al.} 2011, The Astrophysical
  Journal Letters, 740, L13, \dodoi{10.1088/2041-8205/740/1/L13}

\bibitem[{{Gandhi} {et~al.}(2017){Gandhi}, {Bachetti}, {Dhillon}, {Fender},
  {Hardy}, {Harrison}, {Littlefair}, {Malzac}, {Markoff}, {Marsh}, {Mooley},
  {Stern}, {Tomsick}, {Walton}, {Casella}, {Vincentelli}, {Altamirano},
  {Casares}, {Ceccobello}, {Charles}, {Ferrigno}, {Hynes}, {Knigge},
  {Kuulkers}, {Pahari}, {Rahoui}, {Russell}, \& {Shaw}}]{gandhi2017}
{Gandhi}, P., {Bachetti}, M., {Dhillon}, V.~S., {et~al.} 2017, Nature
  Astronomy, 1, 859, \dodoi{10.1038/s41550-017-0273-3}

\bibitem[{Gardner \& Done(2017)}]{gardner2017}
Gardner, E., \& Done, C. 2017, Monthly Notices of the Royal Astronomical
  Society, 470, 3591, \dodoi{10.1093/mnras/stx946}

\bibitem[{{Gierli{\'n}ski} {et~al.}(2009){Gierli{\'n}ski}, {Done}, \&
  {Page}}]{2009MNRAS.392.1106G}
{Gierli{\'n}ski}, M., {Done}, C., \& {Page}, K. 2009, \mnras, 392, 1106,
  \dodoi{10.1111/j.1365-2966.2008.14166.x}

\bibitem[{{Greene} {et~al.}(2001){Greene}, {Bailyn}, \& {Orosz}}]{greene2001}
{Greene}, J., {Bailyn}, C.~D., \& {Orosz}, J.~A. 2001, \apj, 554, 1290,
  \dodoi{10.1086/321411}

\bibitem[{{Haardt} \& {Maraschi}(1991)}]{haardt1991}
{Haardt}, F., \& {Maraschi}, L. 1991, \apjl, 380, L51, \dodoi{10.1086/186171}

\bibitem[{Harmon {et~al.}(1995)Harmon, Wilson, Zhang, Paciesas, Fishman,
  Hjellming, Rupen, Scott, Briggs, \& Rubin}]{harmon1995}
Harmon, B.~A., Wilson, C.~A., Zhang, S.~N., {et~al.} 1995, Nature, 374, 703,
  \dodoi{10.1038/374703a0}

\bibitem[{Hinkle \& Mushotzky(2021)}]{hinkle2021}
Hinkle, J.~T., \& Mushotzky, R. 2021, Monthly Notices of the Royal Astronomical
  Society, 506, 4960, \dodoi{10.1093/mnras/stab1976}

\bibitem[{{Homan} {et~al.}(2005){Homan}, {Buxton}, {Markoff}, {Bailyn},
  {Nespoli}, \& {Belloni}}]{2005ApJ...624..295H}
{Homan}, J., {Buxton}, M., {Markoff}, S., {et~al.} 2005, \apj, 624, 295,
  \dodoi{10.1086/428722}

\bibitem[{Honig {et~al.}(2014)Honig, Watson, Kishimoto, \& Hjorth}]{honig2014}
Honig, S.~F., Watson, D., Kishimoto, M., \& Hjorth, J. 2014, Nature, 515, 528,
  \dodoi{10.1038/nature13914}

\bibitem[{{Horne} {et~al.}(1991){Horne}, {Welsh}, \&
  {Peterson}}]{1991ApJ...367L...5H}
{Horne}, K., {Welsh}, W.~F., \& {Peterson}, B.~M. 1991, \apjl, 367, L5,
  \dodoi{10.1086/185919}

\bibitem[{{Hynes}(2005)}]{2005ASPC..330..237H}
{Hynes}, R.~I. 2005, in Astronomical Society of the Pacific Conference Series,
  Vol. 330, The Astrophysics of Cataclysmic Variables and Related Objects, ed.
  J.~M. {Hameury} \& J.~P. {Lasota}, 237,
  \dodoi{10.48550/arXiv.astro-ph/0410218}

\bibitem[{Hynes {et~al.}(2009)Hynes, Brien, Mullally, \& Ashcraft}]{hynes2009}
Hynes, R.~I., Brien, K.~O., Mullally, F., \& Ashcraft, T. 2009, Monthly Notices
  of the Royal Astronomical Society, 399, 281,
  \dodoi{10.1111/j.1365-2966.2009.15260.x}

\bibitem[{Hynes {et~al.}(1998)Hynes, O'Brien, Horne, Chen, \&
  Haswell}]{hynes1998}
Hynes, R.~I., O'Brien, K., Horne, K., Chen, W., \& Haswell, C.~A. 1998, Monthly
  Notices of the Royal Astronomical Society, 299, L37,
  \dodoi{10.1046/j.1365-8711.1998.01856.x}

\bibitem[{Hynes {et~al.}(2003)Hynes, Haswell, Cui, Shrader, O'Brien, Chaty,
  Skillman, Patterson, \& Horne}]{RN233}
Hynes, R.~I., Haswell, C.~A., Cui, W., {et~al.} 2003, Monthly Notices of the
  Royal Astronomical Society, 345, 292,
  \dodoi{10.1046/j.1365-8711.2003.06938.x}

\bibitem[{Hynes {et~al.}(2019)Hynes, Robinson, Terndrup, Gandhi, Froning,
  Wagner, Starrfield, Dhillon, \& Marsh}]{hynes2019}
Hynes, R.~I., Robinson, E.~L., Terndrup, D.~M., {et~al.} 2019, Monthly Notices
  of the Royal Astronomical Society, 487, 60, \dodoi{10.1093/mnras/stz1238}

\bibitem[{Israel(1967)}]{israel1967}
Israel, W. 1967, Physical Review, 164, 1776, \dodoi{10.1103/PhysRev.164.1776}

\bibitem[{{Jha} {et~al.}(2022){Jha}, {Joshi}, {Chand}, {Wu}, {Ho}, {Rastogi},
  \& {Ma}}]{jha2021}
{Jha}, V.~K., {Joshi}, R., {Chand}, H., {et~al.} 2022, \mnras, 511, 3005,
  \dodoi{10.1093/mnras/stac109}

\bibitem[{Jonker \& Nelemans(2004)}]{RN229}
Jonker, P.~G., \& Nelemans, G. 2004, Monthly Notices of the Royal Astronomical
  Society, 354, 355, \dodoi{10.1111/j.1365-2966.2004.08193.x}

\bibitem[{{Kajava} {et~al.}(2019){Kajava}, {Motta}, {Sanna}, {Veledina}, {Del
  Santo}, \& {Segreto}}]{2019MNRAS.488L..18K}
{Kajava}, J.~J.~E., {Motta}, S.~E., {Sanna}, A., {et~al.} 2019, \mnras, 488,
  L18, \dodoi{10.1093/mnrasl/slz089}

\bibitem[{Kalberla {et~al.}(2005)Kalberla, Burton, Hartmann, Arnal, Bajaja,
  Morras, \& Pöppel}]{kalberla2005}
Kalberla, P. M.~W., Burton, W.~B., Hartmann, D., {et~al.} 2005, Astronomy and
  Astrophysics, 440, 775, \dodoi{10.1051/0004-6361:20041864}

\bibitem[{Kalemci {et~al.}(2013)Kalemci, Dinçer, Tomsick, Buxton, Bailyn, \&
  Chun}]{kalemci2013}
Kalemci, E., Dinçer, T., Tomsick, J.~A., {et~al.} 2013, The Astrophysical
  Journal, 779, 95, \dodoi{10.1088/0004-637X/779/2/95}

\bibitem[{{Kaspi} {et~al.}(2000){Kaspi}, {Smith}, {Netzer}, {Maoz}, {Jannuzi},
  \& {Giveon}}]{2000ApJ...533..631K}
{Kaspi}, S., {Smith}, P.~S., {Netzer}, H., {et~al.} 2000, \apj, 533, 631,
  \dodoi{10.1086/308704}

\bibitem[{Kim {et~al.}(2024)Kim, Di~Gesu, Marin, Marscher, Matt, Soffitta,
  Tombesi, Costa, \& Donnarumma}]{kim2024}
Kim, D.~E., Di~Gesu, L., Marin, F., {et~al.} 2024, Galaxies, 12, 20,
  \dodoi{10.3390/galaxies12030020}

\bibitem[{{Kokubo} \& {Minezaki}(2020)}]{2020MNRAS.491.4615K}
{Kokubo}, M., \& {Minezaki}, T. 2020, \mnras, 491, 4615,
  \dodoi{10.1093/mnras/stz3397}

\bibitem[{Koljonen {et~al.}(2015)Koljonen, Russell, Fernández-Ontiveros,
  Markoff, Russell, Miller-Jones, van~der Horst, Bernardini, Casella, Curran,
  Gandhi, \& Soria}]{koljonen2015}
Koljonen, K. I.~I., Russell, D.~M., Fernández-Ontiveros, J.~A., {et~al.} 2015,
  The Astrophysical Journal, 814, 139, \dodoi{10.1088/0004-637X/814/2/139}

\bibitem[{{Krolik} {et~al.}(1991){Krolik}, {Horne}, {Kallman}, {Malkan},
  {Edelson}, \& {Kriss}}]{1991ApJ...371..541K}
{Krolik}, J.~H., {Horne}, K., {Kallman}, T.~R., {et~al.} 1991, \apj, 371, 541,
  \dodoi{10.1086/169918}

\bibitem[{Körding {et~al.}(2006)Körding, Jester, \& Fender}]{kording2006}
Körding, E.~G., Jester, S., \& Fender, R. 2006, Monthly Notices of the Royal
  Astronomical Society, 372, 1366, \dodoi{10.1111/j.1365-2966.2006.10954.x}

\bibitem[{Lira {et~al.}(1999)Lira, Lawrence, O'Brien, Johnson, Terlevich, \&
  Bannister}]{lira1999}
Lira, P., Lawrence, A., O'Brien, P., {et~al.} 1999, Monthly Notices of the
  Royal Astronomical Society, 305, 109,
  \dodoi{10.1046/j.1365-8711.1999.02388.x}

\bibitem[{Lu \& Wang(2000)}]{lu2000}
Lu, Y., \& Wang, T. 2000, The Astrophysical Journal, 537, L103,
  \dodoi{10.1086/312768}

\bibitem[{{Lusso} \& {Risaliti}(2016)}]{Lusso2016}
{Lusso}, E., \& {Risaliti}, G. 2016, \apj, 819, 154,
  \dodoi{10.3847/0004-637X/819/2/154}

\bibitem[{{Lusso} \& {Risaliti}(2017)}]{Lusso2017}
---. 2017, \aap, 602, A79, \dodoi{10.1051/0004-6361/201630079}

\bibitem[{{Lusso} {et~al.}(2010){Lusso}, {Comastri}, {Vignali}, {Zamorani},
  {Brusa}, {Gilli}, {Iwasawa}, {Salvato}, {Civano}, {Elvis}, {Merloni},
  {Bongiorno}, {Trump}, {Koekemoer}, {Schinnerer}, {Le Floc'h}, {Cappelluti},
  {Jahnke}, {Sargent}, {Silverman}, {Mainieri}, {Fiore}, {Bolzonella}, {Le
  F{\`e}vre}, {Garilli}, {Iovino}, {Kneib}, {Lamareille}, {Lilly}, {Mignoli},
  {Scodeggio}, \& {Vergani}}]{Lusso2010}
{Lusso}, E., {Comastri}, A., {Vignali}, C., {et~al.} 2010, \aap, 512, A34,
  \dodoi{10.1051/0004-6361/200913298}

\bibitem[{{Lyubarskii}(1997)}]{Lyubarskii1997}
{Lyubarskii}, Y.~E. 1997, \mnras, 292, 679, \dodoi{10.1093/mnras/292.3.679}

\bibitem[{{Marcel} {et~al.}(2018{\natexlab{a}}){Marcel}, {Ferreira},
  {Petrucci}, {Henri}, {Belmont}, {Clavel}, {Malzac}, {Coriat}, {Corbel},
  {Rodriguez}, {Loh}, {Chakravorty}, \& {Drappeau}}]{Marcel2018a}
{Marcel}, G., {Ferreira}, J., {Petrucci}, P.-O., {et~al.} 2018{\natexlab{a}},
  \aap, 615, A57, \dodoi{10.1051/0004-6361/201732069}

\bibitem[{{Marcel} {et~al.}(2018{\natexlab{b}}){Marcel}, {Ferreira},
  {Petrucci}, {Belmont}, {Malzac}, {Clavel}, {Henri}, {Coriat}, {Corbel},
  {Rodriguez}, {Loh}, \& {Chakravorty}}]{Marcel2018b}
---. 2018{\natexlab{b}}, \aap, 617, A46, \dodoi{10.1051/0004-6361/201833124}

\bibitem[{{Marcel} {et~al.}(2019){Marcel}, {Ferreira}, {Clavel}, {Petrucci},
  {Malzac}, {Corbel}, {Rodriguez}, {Belmont}, {Coriat}, {Henri}, \&
  {Cangemi}}]{Marcel2019}
{Marcel}, G., {Ferreira}, J., {Clavel}, M., {et~al.} 2019, \aap, 626, A115,
  \dodoi{10.1051/0004-6361/201935060}

\bibitem[{Markert {et~al.}(1973)Markert, Canizares, Clark, Lewin, Schnopper, \&
  Sprott}]{marker1973}
Markert, T.~H., Canizares, C.~R., Clark, G.~W., {et~al.} 1973, The
  Astrophysical Journal, 184, L67, \dodoi{10.1086/181290}

\bibitem[{McHardy {et~al.}(2006)McHardy, Koerding, Knigge, Uttley, \&
  Fender}]{mchardy2006}
McHardy, I.~M., Koerding, E., Knigge, C., Uttley, P., \& Fender, R.~P. 2006,
  Nature, 444, 730, \dodoi{10.1038/nature05389}

\bibitem[{McHardy {et~al.}(2004)McHardy, Papadakis, Uttley, Page, \&
  Mason}]{mchardy2004}
McHardy, I.~M., Papadakis, I.~E., Uttley, P., Page, M.~J., \& Mason, K.~O.
  2004, Monthly Notices of the Royal Astronomical Society, 348, 783,
  \dodoi{10.1111/j.1365-2966.2004.07376.x}

\bibitem[{{McHardy} {et~al.}(2014){McHardy}, {Cameron}, {Dwelly}, {Connolly},
  {Lira}, {Emmanoulopoulos}, {Gelbord}, {Breedt}, {Arevalo}, \&
  {Uttley}}]{mchardy2014}
{McHardy}, I.~M., {Cameron}, D.~T., {Dwelly}, T., {et~al.} 2014, \mnras, 444,
  1469, \dodoi{10.1093/mnras/stu1636}

\bibitem[{McHardy {et~al.}(2016)McHardy, Connolly, Peterson, Bieryla, Chand,
  Elvis, Emmanoulopoulos, Falco, Gandhi, Kaspi, Latham, Lira, McCully, Netzer,
  \& Uemura}]{mchardy2016}
McHardy, I.~M., Connolly, S.~D., Peterson, B.~M., {et~al.} 2016, Astronomische
  Nachrichten, 337, 500, \dodoi{10.1002/asna.201612337}

\bibitem[{McHardy {et~al.}(2018)McHardy, Connolly, Horne, Cackett, Gelbord,
  Peterson, Pahari, Gehrels, Goad, Lira, Arevalo, Baldi, Brandt, Breedt, Chand,
  Dewangan, Done, Elvis, Emmanoulopoulos, Fausnaugh, Kaspi, Kochanek, Korista,
  Papadakis, Rao, Uttley, Vestergaard, \& Ward}]{mchard2018}
McHardy, I.~M., Connolly, S.~D., Horne, K., {et~al.} 2018, Monthly Notices of
  the Royal Astronomical Society, 480, 2881, \dodoi{10.1093/mnras/sty1983}

\bibitem[{{McHardy} {et~al.}(2023){McHardy}, {Beard}, {Breedt}, {Knapen},
  {Vincentelli}, {Veresvarska}, {Dhillon}, {Marsh}, {Littlefair}, {Horne},
  {Glew}, {Goad}, {Kammoun}, \& {Emmanoulopoulos}}]{McHardy2023}
{McHardy}, I.~M., {Beard}, M., {Breedt}, E., {et~al.} 2023, \mnras, 519, 3366,
  \dodoi{10.1093/mnras/stac3651}

\bibitem[{Merloni {et~al.}(2003)Merloni, Heinz, \& Di~Matteo}]{merloni2003}
Merloni, A., Heinz, S., \& Di~Matteo, T. 2003, Monthly Notices of the Royal
  Astronomical Society, 345, 1057, \dodoi{10.1046/j.1365-2966.2003.07017.x}

\bibitem[{Migliari {et~al.}(2007)Migliari, Tomsick, Markoff, Kalemci, Bailyn,
  Buxton, Corbel, Fender, \& Kaaret}]{RN68}
Migliari, S., Tomsick, J.~A., Markoff, S., {et~al.} 2007, The Astrophysical
  Journal, 670, 610, \dodoi{10.1086/522023}

\bibitem[{{Narayan} {et~al.}(1998){Narayan}, {Mahadevan}, \&
  {Quataert}}]{abramowicz1998theory2}
{Narayan}, R., {Mahadevan}, R., \& {Quataert}, E. 1998, in Theory of Black Hole
  Accretion Disks, ed. M.~A. {Abramowicz}, G.~{Bj{\"o}rnsson}, \& J.~E.
  {Pringle}, 148--182, \dodoi{10.48550/arXiv.astro-ph/9803141}

\bibitem[{Netzer(2015)}]{netzer2015}
Netzer, H. 2015, Annual Review of Astronomy and Astrophysics, 53, 365,
  \dodoi{10.1146/annurev-astro-082214-122302}

\bibitem[{Noda {et~al.}(2016)Noda, Minezaki, Watanabe, Kokubo, Kawaguchi, Itoh,
  Morihana, Saito, Nakao, Imai, Moritani, Takaki, Kawabata, Nakaoka, Uemura,
  Kawabata, Yoshida, Arai, Takagi, Morokuma, Doi, Itoh, Yamada, Nakazawa,
  Fukazawa, \& Makishima}]{noda2016}
Noda, H., Minezaki, T., Watanabe, M., {et~al.} 2016, The Astrophysical Journal,
  828, \dodoi{10.3847/0004-637x/828/2/78}

\bibitem[{{O'Brien} {et~al.}(2002){O'Brien}, {Horne}, {Hynes}, {Chen},
  {Haswell}, \& {Still}}]{obrien2002}
{O'Brien}, K., {Horne}, K., {Hynes}, R.~I., {et~al.} 2002, \mnras, 334, 426,
  \dodoi{10.1046/j.1365-8711.2002.05530.x}

\bibitem[{{Orosz} {et~al.}(1997){Orosz}, {Remillard}, {Bailyn}, \&
  {McClintock}}]{Orosz1997}
{Orosz}, J.~A., {Remillard}, R.~A., {Bailyn}, C.~D., \& {McClintock}, J.~E.
  1997, \apjl, 478, L83, \dodoi{10.1086/310553}

\bibitem[{{Paice} {et~al.}(2019){Paice}, {Gandhi}, {Shahbaz}, {Uttley},
  {Arzoumanian}, {Charles}, {Dhillon}, {Gendreau}, {Littlefair}, {Malzac},
  {Markoff}, {Marsh}, {Misra}, {Russell}, \& {Veledina}}]{paice2019}
{Paice}, J.~A., {Gandhi}, P., {Shahbaz}, T., {et~al.} 2019, \mnras, 490, L62,
  \dodoi{10.1093/mnrasl/slz148}

\bibitem[{Palmer {et~al.}(2005)Palmer, Barthelmey, Cummings, Gehrels, Krimm,
  Markwardt, Sakamoto, \& Tueller}]{palmer2005}
Palmer, D.~M., Barthelmey, S.~D., Cummings, J.~R., {et~al.} 2005, The
  Astronomer's Telegram, 546, 1.
\newblock \url{https://ui.adsabs.harvard.edu/abs/2005ATel..546....1P}

\bibitem[{{Pancoast} {et~al.}(2014){Pancoast}, {Brewer}, {Treu}, {Park},
  {Barth}, {Bentz}, \& {Woo}}]{Pancoast2014}
{Pancoast}, A., {Brewer}, B.~J., {Treu}, T., {et~al.} 2014, \mnras, 445, 3073,
  \dodoi{10.1093/mnras/stu1419}

\bibitem[{Peterson {et~al.}(2005)Peterson, Bentz, Desroches, Filippenko, Ho,
  Kaspi, Laor, Maoz, Moran, Pogge, \& Quillen}]{peters2005}
Peterson, B.~M., Bentz, M.~C., Desroches, L.-B., {et~al.} 2005, The
  Astrophysical Journal, 632, 799, \dodoi{10.1086/444494}

\bibitem[{{Plotkin} {et~al.}(2012){Plotkin}, {Markoff}, {Kelly}, {K{\"o}rding},
  \& {Anderson}}]{Plotkin2012}
{Plotkin}, R.~M., {Markoff}, S., {Kelly}, B.~C., {K{\"o}rding}, E., \&
  {Anderson}, S.~F. 2012, \mnras, 419, 267,
  \dodoi{10.1111/j.1365-2966.2011.19689.x}

\bibitem[{Plotkin {et~al.}(2016)Plotkin, Miller-Jones, Jonker, Homan, Russell,
  \& Tomsick}]{plotki2016}
Plotkin, R.~M., Miller-Jones, J. C.~A., Jonker, P.~G., {et~al.} 2016, The
  Astronomer's Telegram, 9765, 1.
\newblock \url{https://ui.adsabs.harvard.edu/abs/2016ATel.9765....1P}

\bibitem[{{Poutanen}(1998)}]{abramowicz1998theory}
{Poutanen}, J. 1998, in Theory of Black Hole Accretion Disks, ed. M.~A.
  {Abramowicz}, G.~{Bj{\"o}rnsson}, \& J.~E. {Pringle}, 100--122,
  \dodoi{10.48550/arXiv.astro-ph/9805025}

\bibitem[{Remillard {et~al.}(2000)Remillard, Morgan, Smith, \&
  Smith}]{remill2000}
Remillard, R., Morgan, E., Smith, D., \& Smith, E. 2000, International
  Astronomical Union Circular, 7389, 2.
\newblock \url{https://ui.adsabs.harvard.edu/abs/2000IAUC.7389....2R}

\bibitem[{{Remillard} \& {McClintock}(2006)}]{Remillard2006}
{Remillard}, R.~A., \& {McClintock}, J.~E. 2006, \araa, 44, 49,
  \dodoi{10.1146/annurev.astro.44.051905.092532}

\bibitem[{{Risaliti} \& {Lusso}(2019)}]{Risaliti2019}
{Risaliti}, G., \& {Lusso}, E. 2019, Nature Astronomy, 3, 272,
  \dodoi{10.1038/s41550-018-0657-z}

\bibitem[{Russell {et~al.}(2006)Russell, Fender, Hynes, Brocksopp, Homan,
  Jonker, \& Buxton}]{RN6}
Russell, D.~M., Fender, R.~P., Hynes, R.~I., {et~al.} 2006, Monthly Notices of
  the Royal Astronomical Society, 371, 1334,
  \dodoi{10.1111/j.1365-2966.2006.10756.x}

\bibitem[{{Russell} {et~al.}(2010){Russell}, {Lewis}, {Roche}, {Clark},
  {Breedt}, \& {Fender}}]{Russell2010}
{Russell}, D.~M., {Lewis}, F., {Roche}, P., {et~al.} 2010, \mnras, 402, 2671,
  \dodoi{10.1111/j.1365-2966.2009.16098.x}

\bibitem[{{Russell} {et~al.}(2007){Russell}, {Maccarone}, {K{\"o}rding}, \&
  {Homan}}]{Russell2007}
{Russell}, D.~M., {Maccarone}, T.~J., {K{\"o}rding}, E.~G., \& {Homan}, J.
  2007, \mnras, 379, 1401, \dodoi{10.1111/j.1365-2966.2007.11996.x}

\bibitem[{{Russell} {et~al.}(2019){Russell}, {Bramich}, {Lewis}, {AlMannaei},
  {Al Qaissieh}, {Al Qasim}, {Al Yazeedi}, {Baglio}, {Bernardini}, {Elgalad},
  {Gabuya}, {Lasota}, {Palado}, {Roche}, {Shivkumar}, {Udrescu}, \&
  {Zhang}}]{2019AN....340..278R}
{Russell}, D.~M., {Bramich}, D.~M., {Lewis}, F., {et~al.} 2019, Astronomische
  Nachrichten, 340, 278, \dodoi{10.1002/asna.201913610}

\bibitem[{{Sacchi} {et~al.}(2025){Sacchi}, {Risaliti}, {Signorini}, {Nardini},
  {Lusso}, \& {Trefoloni}}]{Sacchi2025}
{Sacchi}, A., {Risaliti}, G., {Signorini}, M., {et~al.} 2025, \aap, 703, A273,
  \dodoi{10.1051/0004-6361/202556708}

\bibitem[{{Saikia} {et~al.}(2018){Saikia}, {K{\"o}rding}, {Coppejans},
  {Falcke}, {Williams}, {Baldi}, {Mchardy}, \& {Beswick}}]{2018A&A...616A.152S}
{Saikia}, P., {K{\"o}rding}, E., {Coppejans}, D.~L., {et~al.} 2018, \aap, 616,
  A152, \dodoi{10.1051/0004-6361/201833233}

\bibitem[{Saikia {et~al.}(2015)Saikia, Körding, \& Falcke}]{saikia2015}
Saikia, P., Körding, E., \& Falcke, H. 2015, Monthly Notices of the Royal
  Astronomical Society, 450, 2317, \dodoi{10.1093/mnras/stv731}

\bibitem[{Saikia {et~al.}(2019)Saikia, Russell, Bramich, Miller-Jones, Baglio,
  \& Degenaar}]{RN225}
Saikia, P., Russell, D.~M., Bramich, D.~M., {et~al.} 2019, Astrophysical
  Journal, 887, 21, \dodoi{10.3847/1538-4357/ab4a09}

\bibitem[{{Saikia} {et~al.}(2022){Saikia}, {Russell}, {Baglio}, {Bramich},
  {Casella}, {Diaz Trigo}, {Gandhi}, {Jiang}, {Maccarone}, {Soria}, {Al Noori},
  {Al Yazeedi}, {Alabarta}, {Belloni}, {Bel}, {Ceccobello}, {Corbel}, {Fender},
  {Gallo}, {Homan}, {Koljonen}, {Lewis}, {Markoff}, {Miller-Jones},
  {Rodriguez}, {Russell}, {Shahbaz}, {Sivakoff}, {Testa}, \&
  {Tetarenko}}]{Saikia2022}
{Saikia}, P., {Russell}, D.~M., {Baglio}, M.~C., {et~al.} 2022, \apj, 932, 38,
  \dodoi{10.3847/1538-4357/ac6ce1}

\bibitem[{{Saikia} {et~al.}(2026){Saikia}, {Russell}, {Bramich}, {Alabarta},
  {Rout}, {Vincentelli}, {Mendez}, {Altamirano}, {Garcia}, {Baglio}, {Lewis},
  \& {Yang}}]{Saikia2026}
{Saikia}, P., {Russell}, D.~M., {Bramich}, D.~M., {et~al.} 2026, \mnras, 546,
  stag138, \dodoi{10.1093/mnras/stag138}

\bibitem[{Schlafly \& Finkbeiner(2011)}]{schlaf2011}
Schlafly, E.~F., \& Finkbeiner, D.~P. 2011, The Astrophysical Journal, 737,
  103, \dodoi{10.1088/0004-637x/737/2/103}

\bibitem[{Shahbaz {et~al.}(2015)Shahbaz, Linares, Nevado, Rodríguez-Gil,
  Casares, Dhillon, Marsh, Littlefair, Leckngam, \& Poshyachinda}]{shabhaz2015}
Shahbaz, T., Linares, M., Nevado, S.~P., {et~al.} 2015, Monthly Notices of the
  Royal Astronomical Society, 453, 3461, \dodoi{10.1093/mnras/stv1686}

\bibitem[{{Shakura} \& {Sunyaev}(1973)}]{ss73}
{Shakura}, N.~I., \& {Sunyaev}, R.~A. 1973, \aap, 24, 337

\bibitem[{Shappee {et~al.}(2014)Shappee, Prieto, Grupe, Kochanek, Stanek,
  De~Rosa, Mathur, Zu, Peterson, Pogge, Komossa, Im, Jencson, Holoien, Basu,
  Beacom, Szczygieł, Brimacombe, Adams, Campillay, Choi, Contreras, Dietrich,
  Dubberley, Elphick, Foale, Giustini, Gonzalez, Hawkins, Howell, Hsiao, Koss,
  Leighly, Morrell, Mudd, Mullins, Nugent, Parrent, Phillips, Pojmanski,
  Rosing, Ross, Sand, Terndrup, Valenti, Walker, \& Yoon}]{shappee2014}
Shappee, B.~J., Prieto, J.~L., Grupe, D., {et~al.} 2014, The Astrophysical
  Journal, 788, 48, \dodoi{10.1088/0004-637X/788/1/48}

\bibitem[{Shaw {et~al.}(2018)Shaw, Tetarenko, Dubus, Dinçer, Tomsick, Gandhi,
  Plotkin, \& Russell}]{shaw2018}
Shaw, A.~W., Tetarenko, B.~E., Dubus, G., {et~al.} 2018, Monthly Notices of the
  Royal Astronomical Society, 482, 1840, \dodoi{10.1093/mnras/sty2787}

\bibitem[{Steele {et~al.}(2004)Steele, Smith, Rees, Baker, Bates, Bode, Bowman,
  Carter, Etherton, Ford, Fraser, Gomboc, Lett, Mansfield, Marchant,
  Medrano-Cerda, Mottram, Raback, Scott, Tomlinson, \& Zamanov}]{liverpool}
Steele, I.~A., Smith, R.~J., Rees, P.~C., {et~al.} 2004, in Ground-based
  Telescopes, ed. J.~Oschmann, Jacobus~M., Vol. 5489, 679--692,
  \dodoi{10.1117/12.551456}

\bibitem[{{Strateva} {et~al.}(2005){Strateva}, {Brandt}, {Schneider}, {Vanden
  Berk}, \& {Vignali}}]{Strateva2005}
{Strateva}, I.~V., {Brandt}, W.~N., {Schneider}, D.~P., {Vanden Berk}, D.~G.,
  \& {Vignali}, C. 2005, \aj, 130, 387, \dodoi{10.1086/431247}

\bibitem[{{Subasavage} {et~al.}(2010){Subasavage}, {Bailyn}, {Smith}, {Henry},
  {Walter}, \& {Buxton}}]{Subasavage2010}
{Subasavage}, J.~P., {Bailyn}, C.~D., {Smith}, R.~C., {et~al.} 2010, in Society
  of Photo-Optical Instrumentation Engineers (SPIE) Conference Series, Vol.
  7737, Observatory Operations: Strategies, Processes, and Systems III, ed.
  D.~R. {Silva}, A.~B. {Peck}, \& B.~T. {Soifer}, 77371C,
  \dodoi{10.1117/12.859145}

\bibitem[{{Suganuma} {et~al.}(2006){Suganuma}, {Yoshii}, {Kobayashi},
  {Minezaki}, {Enya}, {Tomita}, {Aoki}, {Koshida}, \&
  {Peterson}}]{2006ApJ...639...46S}
{Suganuma}, M., {Yoshii}, Y., {Kobayashi}, Y., {et~al.} 2006, \apj, 639, 46,
  \dodoi{10.1086/499326}

\bibitem[{{Tetarenko} {et~al.}(2021){Tetarenko}, {Casella}, {Miller-Jones},
  {Sivakoff}, {Paice}, {Vincentelli}, {Maccarone}, {Gandhi}, {Dhillon},
  {Marsh}, {Russell}, \& {Uttley}}]{Tetarenko2021}
{Tetarenko}, A.~J., {Casella}, P., {Miller-Jones}, J.~C.~A., {et~al.} 2021,
  \mnras, 504, 3862, \dodoi{10.1093/mnras/stab820}

\bibitem[{Thim {et~al.}(2004)Thim, Hoessel, Saha, Claver, Dolphin, \&
  Tammann}]{thim2004}
Thim, F., Hoessel, J.~G., Saha, A., {et~al.} 2004, The Astronomical Journal,
  127, 2322, \dodoi{10.1086/382244}

\bibitem[{Tody(1986)}]{iraf1986}
Tody, D. 1986, in Instrumentatidon in astronomy VI, ed. D.~L. Crawford, Vol.
  627, 733, \dodoi{10.1117/12.968154}

\bibitem[{Tody(1993)}]{iraf1993}
Tody, D. 1993, in Astronomical Data Analysis Software and Systems II, ed. R.~J.
  Hanisch, R.~J.~V. Brissenden, \& J.~Barnes, Vol.~52, 173.
\newblock \url{https://ui.adsabs.harvard.edu/abs/1993ASPC...52..173T}

\bibitem[{Tomsick {et~al.}(2015)Tomsick, Rahoui, Kolehmainen, Miller-Jones,
  Fürst, Yamaoka, Akitaya, Corbel, Coriat, Done, Gandhi, Harrison, Huang,
  Kaaret, Kalemci, Kanda, Migliari, Miller, Moritani, Stern, Uemura, \&
  Urata}]{tomsic2015}
Tomsick, J.~A., Rahoui, F., Kolehmainen, M., {et~al.} 2015, The Astrophysical
  Journal, 808, 85, \dodoi{10.1088/0004-637X/808/1/85}

\bibitem[{{Torres} {et~al.}(2020){Torres}, {Casares}, {Jim{\'e}nez-Ibarra},
  {{\'A}lvarez-Hern{\'a}ndez}, {Mu{\~n}oz-Darias}, {Armas Padilla}, {Jonker},
  \& {Heida}}]{2020ApJ...893L..37T}
{Torres}, M.~A.~P., {Casares}, J., {Jim{\'e}nez-Ibarra}, F., {et~al.} 2020,
  \apjl, 893, L37, \dodoi{10.3847/2041-8213/ab863a}

\bibitem[{{Torres} {et~al.}(2019){Torres}, {Casares}, {Jim{\'e}nez-Ibarra},
  {Mu{\~n}oz-Darias}, {Armas Padilla}, {Jonker}, \&
  {Heida}}]{2019ApJ...882L..21T}
---. 2019, \apjl, 882, L21, \dodoi{10.3847/2041-8213/ab39df}

\bibitem[{Troyer {et~al.}(2016)Troyer, Starkey, Cackett, Bentz, Goad, Horne, \&
  Seals}]{troyer2016}
Troyer, J., Starkey, D., Cackett, E.~M., {et~al.} 2016, Monthly Notices of the
  Royal Astronomical Society, 456, 4040, \dodoi{10.1093/mnras/stv2862}

\bibitem[{{Tucker} {et~al.}(2018){Tucker}, {Shappee}, {Holoien}, {Auchettl},
  {Strader}, {Stanek}, {Kochanek}, {Bahramian}, {ASAS-SN}, {Dong}, {Prieto},
  {Shields}, {Thompson}, {Beacom}, {Chomiuk}, {ATLAS}, {Denneau}, {Flewelling},
  {Heinze}, {Smith}, {Stalder}, {Tonry}, {Weiland}, {Rest}, {Huber}, {Rowan},
  \& {Dage}}]{2018ApJ...867L...9T}
{Tucker}, M.~A., {Shappee}, B.~J., {Holoien}, T.~W.~S., {et~al.} 2018, \apjl,
  867, L9, \dodoi{10.3847/2041-8213/aae88a}

\bibitem[{Turner {et~al.}(2017)Turner, Miller, Reeves, \& Braito}]{turner2017}
Turner, T.~J., Miller, L., Reeves, J.~N., \& Braito, V. 2017, Monthly Notices
  of the Royal Astronomical Society, 467, 3924, \dodoi{10.1093/mnras/stx388}

\bibitem[{{Uttley} {et~al.}(2003){Uttley}, {Edelson}, {McHardy}, {Peterson}, \&
  {Markowitz}}]{2003ApJ...584L..53U}
{Uttley}, P., {Edelson}, R., {McHardy}, I.~M., {Peterson}, B.~M., \&
  {Markowitz}, A. 2003, \apjl, 584, L53, \dodoi{10.1086/373887}

\bibitem[{{Uttley} {et~al.}(2011){Uttley}, {Wilkinson}, {Cassatella}, {Wilms},
  {Pottschmidt}, {Hanke}, \& {B{\"o}ck}}]{2011MNRAS.414L..60U}
{Uttley}, P., {Wilkinson}, T., {Cassatella}, P., {et~al.} 2011, \mnras, 414,
  L60, \dodoi{10.1111/j.1745-3933.2011.01056.x}

\bibitem[{{van Paradijs} \& {McClintock}(1994)}]{RN5}
{van Paradijs}, J., \& {McClintock}, J.~E. 1994, \aap, 290, 133

\bibitem[{{Veledina} {et~al.}(2011){Veledina}, {Poutanen}, \&
  {Vurm}}]{veledina2011}
{Veledina}, A., {Poutanen}, J., \& {Vurm}, I. 2011, \apjl, 737, L17,
  \dodoi{10.1088/2041-8205/737/1/L17}

\bibitem[{{Veledina} {et~al.}(2013){Veledina}, {Poutanen}, \&
  {Vurm}}]{veledina2013}
---. 2013, \mnras, 430, 3196, \dodoi{10.1093/mnras/stt124}

\bibitem[{{Vignali} {et~al.}(2003){Vignali}, {Brandt}, \&
  {Schneider}}]{Vignali2003}
{Vignali}, C., {Brandt}, W.~N., \& {Schneider}, D.~P. 2003, \aj, 125, 433,
  \dodoi{10.1086/345973}

\bibitem[{{Vincentelli} {et~al.}(2020){Vincentelli}, {Cavecchi}, {Casella},
  {Migliari}, {Altamirano}, {Belloni}, \& {Diaz-Trigo}}]{2020MNRAS.495L..37V}
{Vincentelli}, F.~M., {Cavecchi}, Y., {Casella}, P., {et~al.} 2020, \mnras,
  495, L37, \dodoi{10.1093/mnrasl/slaa049}

\bibitem[{Walton {et~al.}(2013)Walton, Zoghbi, Cackett, Uttley, Harrison,
  Fabian, Kara, Miller, Reis, \& Reynolds}]{walton2013}
Walton, D.~J., Zoghbi, A., Cackett, E.~M., {et~al.} 2013, The Astrophysical
  Journal Letters, 777, L23, \dodoi{10.1088/2041-8205/777/2/L23}

\bibitem[{Wilkins \& Fabian(2012)}]{wilkins2012}
Wilkins, D.~R., \& Fabian, A.~C. 2012, Monthly Notices of the Royal
  Astronomical Society, 424, 1284, \dodoi{10.1111/j.1365-2966.2012.21308.x}

\bibitem[{Wong {et~al.}(2024)Wong, Medeiros, Cárdenas-Avendaño, \&
  Stone}]{wong2024}
Wong, G.~N., Medeiros, L., Cárdenas-Avendaño, A., \& Stone, J.~M. 2024, The
  Astrophysical Journal Letters, 975, L40, \dodoi{10.3847/2041-8213/ad8650}

\bibitem[{Woo {et~al.}(2019)Woo, Cho, Gallo, Hodges-Kluck, Le, Shin, Son, \&
  Horst}]{woo2019}
Woo, J.-H., Cho, H., Gallo, E., {et~al.} 2019, Nature Astronomy, 3, 755,
  \dodoi{10.1038/s41550-019-0790-3}

\bibitem[{{Wu} \& {Han}(2001)}]{Wu2001}
{Wu}, X.-B., \& {Han}, J.~L. 2001, \apjl, 561, L59, \dodoi{10.1086/324408}

\bibitem[{{Yanes-Rizo} {et~al.}(2025){Yanes-Rizo}, {Torres}, {Casares},
  {Jonker}, {S{\'a}nchez-Sierras}, {Mu{\~n}oz-Darias}, \& {Armas
  Padilla}}]{2025AA...694A.119Y}
{Yanes-Rizo}, I.~V., {Torres}, M.~A.~P., {Casares}, J., {et~al.} 2025, \aap,
  694, A119, \dodoi{10.1051/0004-6361/202452575}

\bibitem[{{Yang} {et~al.}(2022){Yang}, {Zhang}, {Russell}, {Gelfand},
  {M{\'e}ndez}, {Wang}, \& {Lyu}}]{Yang2022}
{Yang}, P., {Zhang}, G., {Russell}, D.~M., {et~al.} 2022, \mnras, 514, 234,
  \dodoi{10.1093/mnras/stac1120}

\bibitem[{{Yang} {et~al.}(2025){Yang}, {Zhang}, {Russell}, {M{\'e}ndez},
  {Baglio}, {Altamirano}, {Yang}, {Saikia}, \& {Alabarta}}]{Yang2025}
---. 2025, \aap, 700, A79, \dodoi{10.1051/0004-6361/202555089}

\bibitem[{Yuan \& Narayan(2014)}]{yuan2014}
Yuan, F., \& Narayan, R. 2014, Annual Review of Astronomy and Astrophysics, 52,
  529, \dodoi{10.1146/annurev-astro-082812-141003}

\bibitem[{Yuan {et~al.}(2021)Yuan, Macri, Peterson, Riess, Fausnaugh, Hoffmann,
  Anand, Bentz, Dalla~Bontà, Davies, De~Rosa, Ferrarese, Grier, Hicks, Onken,
  Pogge, Storchi-Bergmann, \& Vestergaard}]{yuan2021}
Yuan, W., Macri, L.~M., Peterson, B.~M., {et~al.} 2021, The Astrophysical
  Journal, 913, 3, \dodoi{10.3847/1538-4357/abf24a}

\bibitem[{{Zdziarski} {et~al.}(2021){Zdziarski}, {Dzie{\l}ak}, {De Marco},
  {Szanecki}, \& {Nied{\'z}wiecki}}]{2021ApJ...909L...9Z}
{Zdziarski}, A.~A., {Dzie{\l}ak}, M.~A., {De Marco}, B., {Szanecki}, M., \&
  {Nied{\'z}wiecki}, A. 2021, \apjl, 909, L9, \dodoi{10.3847/2041-8213/abe7ef}

\bibitem[{{Zhang} {et~al.}(2019){Zhang}, {Bernardini}, {Russell}, {Gelfand},
  {Lasota}, {Qasim}, {AlMannaei}, {Koljonen}, {Shaw}, {Lewis}, {Tomsick},
  {Plotkin}, {Miller-Jones}, {Maitra}, {Homan}, {Charles}, {Kobel}, {Perez}, \&
  {Doran}}]{2019ApJ...876....5Z}
{Zhang}, G.~B., {Bernardini}, F., {Russell}, D.~M., {et~al.} 2019, \apj, 876,
  5, \dodoi{10.3847/1538-4357/ab12dd}

\end{thebibliography}

\appendix

\section{Removing Host Galaxy Contribution of NGC 4151}
\label{app:host}

To accurately quantify AGN variability and flux from aperture photometry, we subtracted the host galaxy contribution in the visual V band. Photometry was performed using IRAF software \citep{iraf1986, iraf1993}, utilizing a 5 arcsec radius circular aperture with a background region selected to represent the sky near the galaxy.

To estimate the constant contribution from non-nuclear light, we analyzed archival Las Cumbres Observatory (LCO) $V$-band images, benefiting from the higher resolution compared to UVOT. Processed images were retrieved from the LCO Science Archive \footnote{\url{https://archive.lco.global}} and corresponded to the same observation period as our optical flux data. Using IRAF, we determined the point spread function (PSF), masking out bright irregular features and Galactic stars during the fitting process. The sky annulus was set at 1.5 times the size of the host galaxy aperture.

From the LCO image of NGC 4151, we subtracted the AGN's PSF and sky background. The circular aperture and its background annulus, which included some host galaxy light, were duplicated to measure the host galaxy flux. The $V$-band host galaxy flux was determined to be $1.295 \times 10^{-14}$ erg s$^{-1}$ cm$^{-2}$ \AA$^{-1}$.

\end{document}